\documentclass[a4paper,11pt]{article}
\usepackage{jheppub} 
\usepackage{lineno}
\usepackage{amsmath}
\usepackage{tikz}
\usepackage{spectralsequences}
\usepackage{hyperref}
\usepackage{jheppub}
\usepackage{tikz-cd}
\usepackage{pgfplots}
\usepackage{tikz-3dplot}
\usepackage{braket}
\usepackage{tikz,lipsum,lmodern}
\usepackage[most]{tcolorbox}

\usetikzlibrary{matrix}

\title{\boldmath Generalized Instanton Symmetry Induced by Monopoles}

\author{Aiden Sheckler}
\affiliation{Department of Physics, University of California, San Diego,\\
9500 Gilman Drive, La Jolla CA 92093-0319, USA}

\emailAdd{asheckler@ucsd.edu}

\abstract{We point out that there exists a generalization of instanton symmmetry in the Coulomb phase of 5d nonabelian gauge theories which is capable of measuring a wider class of topological charges of monopole strings. The symmetry is invertible on compact spacetimes, but non-invertible on spacetimes with boundary. In the case of maximal supersymmetry, we show that this symmetry has a natural origin coming from the 6d $\mathcal N=(2,0)$ superconformal field theory under dimensional reduction. By generalizing this construction to any ADE gauge group, this allows us to propose a broad new class of homotopy invariants provided by the boundary non-invertible symmetry defects.}

\begin{document}

\maketitle

\section{Introduction}

Symmetry provides a powerful organizing framework in quantum field theory. Despite our limitations in probing non-perturbative sectors, symmetries provide a rare point of leverage for investigating the physics of these regimes. It is useful to organize the spectrum of excitations in terms of their charges under various symmetries, and to organize phases of theories around how symmetries are manifested. This approach has been super-charged in recent years due to the paradigm shift in how we understand symmetries in QFT, leading to the study of \textit{generalized symmetries} \cite{Gaiotto:2014kfa, Cordova:2022ruw, McGreevy:2022oyu, Brennan:2023mmt, Costa:2024wks, Bhardwaj:2023kri}. In particular, the organization of non-perturbative excitations, such as solitons or defects, folds nicely into this framework, since it has been realized that such objects carry charge under these generalized versions of symmetry. \\

A common class of these objects are topologically-protected excitations, which are defined by the topological winding of the fields. Examples of such objects include solitons such as vortices, skyrmions, or instantons, which are protected by their topological winding number under some homotopy class $\pi_k(X)$ of the field space or gauge group. This winding number defines the topological charge under some generalized symmetry, sometimes referred to as a ``magnetic symmetry" \cite{Gaiotto:2014kfa, Sheckler:2025rlk, Bhardwaj:2022dyt, Pace:2023kyi}. The ubiquity of these excitations makes the study of these symmetries and their realizations a promising direction for gaining insight into strongly-coupled physics. \\

It has recently been appreciated that there can exist nontrivial relationships between different species of solitons when they exist simultaneously, which the techniques of generalized symmetries are perfectly equipped to address \cite{Chen:2022cyw, Chen:2023czk, Pace:2023mdo, Sheckler:2025rlk}. In particular, multiple magnetic symmetries generically mix to form \textit{non-invertible symmetries}, which instead of being described by a (higher) group must be formulated in the language of (higher) fusion categories. The goal of this work is to descripe how this can occur in the context of 5-dimensional non-abelian gauge theories, which necessarily support instantons and their associated instanton symmetry. \\ 

Instantons play a central role in our understanding of nonabelian gauge theories in dimensions $d\geq 4$. This is especially true in five dimensions, where instantons are non-perturbative particle-like objects, and the topological instanton number is measured by a current $J_I=\frac{1}{8\pi^2}\text{Tr}(F\wedge F)$ which generates a $U(1)_I$ 0-form symmetry \cite{Seiberg:1996bd}. This topological current is the characteristic class which measures (part of) the homotopy class of $\pi_4(BG) = \pi_3(G)$ associated to the gauge group $G$. \\

We will be particularly interested in the Coulomb phase of these theories, where the gauge symmetry is spontaneously broken $G\to T_G$ by adjoint matter $\Phi$ all the way down to the subgroup $T_G$ generated by the Cartan. In this phase, the theory supports another species of topological excitation which is the 't Hooft-Polyakov monopole. These are topologically-protected objects which are defined by non-trivial winding of the vev of $\Phi$ around the vacuum manifold $\pi_2(G/T_G)$. In five dimensions these monopoles are strings. There is a large body of work which has been dedicated to studying how monopole strings and instantons in 5d contribute to the strongly-coupled sectors and UV completions of nonabelian gauge theory. \\

In this work we point out that the above phenomenon, where the charges of different topological excitations can mix into an enriched symmetry structure, may be realized in the Coulomb phase. In particular, when the theory is considered on a spacetime which contains a compact spatial direction $M_5=M_4\times S^1$, then it has been known for a long time that, in addition to their topological magnetic charge, those monopole strings which wrap the spatial circle can carry charge under the instanton symmetry as well \cite{Lee:1997vp, Lee:1998vu, Lambert:2010iw, Tachikawa:2011ch, Kraan:1998kp, Davies:2000nw, Hanany:2001iy}. We will show that the result is that there exists a generalization of the ordinary instanton symmetry which is capable of measuring a wider class of topological charges of these wrapped monopole strings. This case is more subtle than previous incarnations of this phenomenon, since the manifold $G/T_G$ which carries the nontrivial topology that induces this mixing between homotopy groups is neither the field space nor the gauge group. As a result, we will find that the topological symmetry defect operators which generate this generalized instanton symmetry are actually invertible on compact spacetimes, but non-invertible on spacetimes with boundary, with the induced boundary TQFT playing the role of the generalized homotopy invariants studied in \cite{Chen:2022cyw}. We will restrict the bulk of our discussion to the simplest case of $G = SU(2)$, so that the SSB pattern is simply $SU(2)\to U(1)$, however in \hyperref[sec5]{Section 5} we will discuss possible generalizations to arbitrary ADE gauge group. \\ 

After constructing this symmetry, we search for a better understanding of its origin. A convenient setting for this is in the case of maximal supersymmetry, which in 5d is the $\mathcal N=2$ theory. This case is nice for several reasons, such as the fact that this theory has an elegant UV completion as the circle reduction of the 6d $\mathcal N=(2,0)$ theory. This gives the instanton symmetry a manifest interpretation as the Kaluza-Klein symmetry resulting from the compactification. Moreover, the maximal supersymmetry protects the 5d Coulomb branch from quantum corrections, so that it can be mapped onto the 6d tensor branch and the field contents can be identified. However the primary tool we will leverage in this analysis is the Symmetry Topological Field Theory (SymTFT) \cite{Freed:2022qnc, Bhardwaj:2023ayw, Kapustin:2014gua, Ji:2019jhk, Apruzzi:2021nmk, Gaiotto:2020iye}. This is a powerful framework of generalized symmetries which allows for a systematic study of the symmetries of a theory by realizing it as the boundary of a bulk $(d+1)$-dim topological field theory which encodes all aspects of the symmetries, anomalies, and charges. The fact that the 6d $\mathcal N=(2,0)$ theory has a natural formulation as the relative boundary theory of a 7d anomaly topological field theory \cite{Witten:1996hc, Witten:1998wy, Gukov:2020btk} allows for the direct derivation of the SymTFT for the 5d $\mathcal N=2$ theory under dimensional reduction. We will find that through this process then the generalized instanton symmetry is naturally identified. \\

The perspective of the 6d (2,0) theory will also provide us with a pathway to generalizing the study of this refined instanton symmetry to arbitrary ADE gauge group. While we do not undertake an exhaustive study of this generalization in this work, in \hyperref[sec5]{Section 5} we speculate on a possible construction for the symmetry defect operators in the case of general gauge group. Since the non-invertible boundary TQFTs of the symmetry defect operators in the $G=SU(2)$ case can be identified with homotopy invariants which can measure the general topological winding number of the monopole strings, we find it tempting to conjecture that the corresponding boundary TQFTs of these more general symmetry defects play a similar role. \\ 

The rest of this paper is organized as follows. In \hyperref[sec2]{Section 2} we review instantons and 't Hooft-Polyakov monopole strings in 5d nonabelian gauge theories, and argue that the monopole strings can carry a more refined topological winding number which can be identified as a version of instanton number. In \hyperref[sec3]{Section 3} we construct the topological symmetry defect operators in the case of $G=SU(2)$, which are capable of measuring this general topological charge, and see that they indeed look like a torsion-reduced version of the instanton symmetry operators. In \hyperref[sec4]{Section 4} we look for an explanation of the origin of this generalized instanton symmetry in the case of maximal supersymmetry. After reviewing the role of the 6d (2,0) theory as a UV completion, we derive the SymTFT for the 5d theory from the reduction of the 6d (2,0) anomaly theory. From studying the anomalies of this SymTFT, we see that there is a natural candidate for the origin of the generalized instanton symmetry defects. In \hyperref[sec5]{Section 5} we propose a possible generalization to arbitrary ADE gauge group using the perspective provided by the 6d (2,0) theory, and construct the associated symmetry defect operators. Finally in \hyperref[sec6]{Section 6} we discuss future directions and questions, including possible connections to the conjecture that 5d instantons of $SU(N)$ gauge theory are composed of partons. \\

 \textbf{Note:} While the final draft of this work was being prepared the paper \cite{Bertolini:2025wyj} was released, which constructs and analyzes the same SymTFT that we work with in \hyperref[sec4]{Section 4}. However the goals and results of this work are independent from those discussed here, where they primarily focus on the global symmetries of $SU(N)/\mathbb{Z}_k$ 5d gauge theory in the symmetric phase with different global forms of the gauge group.

\section{Instantons and Monopoles in 5d}\label{sec2}

We are interested in the solitons of nonabelian gauge theories in 5d. Our example of principal interest will be $SU(2)$ Yang-Mills with adjoint scalar matter, although we allow for the possibility of additional fields, such as in the case of 5d $\mathcal N=2$ super-Yang-Mills. We will first review these solitonic excitations by uplifting them from 4d into a flat 5d space. We will then consider how the possibilities are modified when one of the spatial directions is a compact $S^1$. \\

\subsection{Flat Space $\mathbb{R}^{1+4}$}

A soliton common to all nonabelian gauge theories in $d\geq 4$ is the instanton. In 5d, these are particle-like excitations. If we consider a factorized spacetime $M_5=M_4\times \mathbb R$, where $M_4$ describes a spatial slice, then an instanton is defined by a gauge field configuration which has nontrivial Pontryagin number on $M_4$, so that:
\begin{equation}
    \frac{1}{8\pi^2}\int_{M_4}\text{Tr}(G\wedge G)=\nu\in \mathbb Z
\end{equation}
where $G_{\mu\nu}$ is the nonabelian field strength. The Pontryagin density $J_I = \frac{1}{8\pi^2}\text{Tr}(G\wedge G)$ acts as a conserved Noether current, and so the instanton number $\nu$ can be thought of as the charge of this solitonic configuration under a $U(1)_I$ symmetry referred to as \textit{instanton symmetry} \cite{Seiberg:1996bd, BenettiGenolini:2020doj, Garcia-Valdecasas:2024cqn, Heidenreich:2020pkc, Lambert:2014jna}. From the perspective of generalized symmetries the fact that the instanton current $J_I$ is closed implies that this symmetry is generated by topological codimension-1 symmetry defect operators $U_{\alpha}(\Sigma_4)=\exp(i\alpha\int_{\Sigma_4}J_I)$. These are supported on some submanifold $\Sigma_4$ which can link the instanton and measure its charge by evaluating the Pontryagin flux. \\

We can also consider the low-energy regime of the theory, where we assume that the adjoint Higgs field $\Phi$ spontaneously breaks the gauge symmetry to its Cartan $SU(2) \to U(1)$ by acquiring a vev $\braket{\Phi}=a\sigma^3$. In this case then another solitonic exciation which can appear is the \textit{'t Hooft-Polyakov monopole}. These are defined by the boundary conditions that the theory must lie in its vacuum at asymptotic infinity, but gauge freedom allows $\Phi(x)$ to wind nontrivially around the vacuum manifold $SU(2)/U(1)$ at spatial infinity. There exist solutions for these monopole configurations which saturate the Bogomol'nyi bound and thus satisfy the BPS equations:
\begin{equation}
    B_i=\pm D_i\Phi
\end{equation}
In 4d such solutions are regarded as static configuration, where the winding is defined around the asymptotic boundary of a spatial slice $\partial \mathbb R^3=S^2_{\infty}$. Such configurations are topologically protected since the map $\Phi : S_{\infty}^2\to SU(2)/U(1)\cong S^2$ has nontrivial homotopy class in $\pi_2(S^2)$. We can embed these solutions in 5d as static line configurations which extend along the additional fifth coordinate without functionally depending on it. Therefore in 5d 't Hooft-Polyakov monopoles are promoted to monopole strings. \\

Let us recall how the solitonic charges of 't Hooft-Polyakov monopoles are measured in the IR after the gauge symmetry has been spontaneously broken. The familiar BPS solution in 4d for the adjoint scalar field is $\Phi^i=a\hat r^if(r)$ where $f(r)\to 1$ as $r\to \infty$ and $f(r)\to 0$ as $r\to 0$. These configurations can be transformed back to the Higgs vacuum $\Phi=a\sigma^3$ by a continuous gauge transformation except for a nontrivial chart boundary surrounding infinity along $S^2_{\infty}$, which imbues the configuration with magnetic charge. The contribution to the total energy from just the scalar is divergent, so in order to define a finite-energy state a compensating gauge field configuration is required to cancel this. The condition for this cancellation is that asymptotically $|D_{\mu}\Phi|^2\to 0$. \\

To see the magnetic charge of this configuration explicitly, consider the surviving component of the gauge field after the $SU(2)\to U(1)$ gauge symmetry breaking, which points in the direction of $\Phi$. The associated field strength is given by the \textit{'t Hooft tensor} \cite{HOOFT1974276}:
\begin{equation}
    \mathcal{F}_{\mu\nu}=\frac 1a\Phi^iG^i_{\mu\nu}-\frac{1}{a^3}\epsilon_{ijk}\Phi^iD_{\mu}\Phi^jD_{\nu}\Phi^k
\end{equation}
which is manifestly gauge invariant. The above finite-energy condition can be used to show that the nonabelian field strength asymptotes to $G^i_{\mu\nu}\to \Phi^i\mathcal F_{\mu\nu}$ at infinity. One can also readily check that $\mathcal F$ can be rewritten as:
\begin{equation}
    \mathcal{F}=\frac 1a d(\Phi^iA^i)-\frac{1}{a^3}\text{Tr}(\Phi d\Phi \wedge d\Phi)
\end{equation}
We recognize the latter term as being equivalent to the pullback $\Phi^*\omega_2$ of the volume form $\omega_2$ of $SU(2)/U(1) \cong S^2$ by $\Phi$. Therefore $\mathcal F$ is also manifestly closed $d\mathcal F=0$. Since $\int_{S^2_{\infty}}\Phi^*\omega_2$ is a homotopy invariant which computes the winding number of $\Phi : S^2_{\infty}\to SU(2)/U(1)$ then we see that a monopole with nontrivial winding number $n\in \pi_2(SU(2)/U(1))$ carries a magnetic charge of $n$. In 5d the constant monopole string therefore carries magnetic charge at asymptotic infinity around $S^2_{\infty}\times \mathbb  R$. \\

If we consider flat space $\mathbb R^{1+4}$ then one might wonder if there are particle-like solitons which arise from nontrivial winding of the Higgs vev around the boundary of a spatial slice $\partial \mathbb R^4\cong S^3_{\infty}$ since there is a nontrivial homotopy group $\pi_3(SU(2)/U(1))$. As we will review, these defects should be identified with small instantons. One can already see that this must be the case since the long exact sequence in homotopy identifies $\pi_3(SU(2)/U(1))\cong \pi_3(SU(2))$. The adjective ``small" refers to the fact that the energetic preference of their radius is sent to zero in the IR. Dyonic solutions can be stabilized by giving them electric charge \cite{Lambert:1999ua}. \\

Indeed, consider the abelian instanton symmetry in the IR, which is measured by the abelian instanton current $\frac{1}{8\pi^2}\int \mathcal F\wedge \mathcal F$. Let us briefly recall how the instanton number is computed on the asymptotic compactification $\mathbb R^4\cup \{\infty \}\cong S^4$. One divides the space into disk charts $D_+\cup_{S^3_{\infty}}D_-$, one of which includes only a small neighborhood surrounding the north pole at infinity, so that they are glued along a boundary given by the 3-sphere at infinity $S^3_{\infty}=\partial \mathbb R^4$. Since the disks are topologically trivial then within each chart we can trivialize the instanton current in terms of the abelian Chern-Simons 3-form $\mathcal F\wedge \mathcal F=dCS(\mathcal F\wedge \mathcal F)$. The total instanton number can then be computed as:
\begin{equation}
    \frac{1}{8\pi^2}\int_{S^4}\mathcal F\wedge \mathcal F=\frac{1}{8\pi^2}\int_{S^3_{\infty,+}-S^3_{\infty,-}}CS(\mathcal F\wedge \mathcal F)
\end{equation}
Thus in the solutions where we impose the winding boundary conditions at asymptotic infinity, then the instanton number is only dependent on the behavior of the field strength at asymptotic infinity. Now we can use the asymptotic form of the nonabelian field strength to see that at infinity the UV nonabelian instanton number becomes $\text{Tr}(G\wedge G)\to \mathcal F\wedge \mathcal F$. Therefore we conclude that the $U(1)_I$ instanton symmetry of the UV manifests in the IR as the abelian instanton symmetry, which is consistent with the analysis of \cite{GarciaGarcia:2025uub}. Notice also that since $CS(\Phi^*\omega_2\wedge \Phi^*\omega_2)$ is the Hopf invariant, which measures the $\pi_3$ winding number of $\Phi : S^3\to SU(2)/U(1)$, then indeed the solitonic configurations of $\Phi$ which wind asymptotically around $S^3_{\infty}\cong\partial \mathbb R^4$ can be identified as instantons. \\

\subsection{Compact Spatial Circle $\mathbb R^{1+3}\times S^1$}

A key insight of this work is that if we consider a 5d theory compactified on a spatial circle, so that the spacetime is $\mathbb R^{1+3}\times S^1$, then the asymptotic boundary of a spatial slice has topology of $S^2_{\infty}\times S^1$. This now allows for monopole lines which can wrap the spatial $S^1$. This will lead to a broader set of possible topological configurations since the set of homotopy classes from $S_{\infty}^2\times S^1$ into $SU(2)/U(1)$ is more complicated than what can be captured by a single homotopy group alone. Indeed, such homotopy classes are classified by \cite{Chen:2022cyw}:
\begin{equation}\label{deforms}
    [S^2\times S^1,S^2]=\{(n,l) : n\in \mathbb Z, \ l\in \mathbb Z_{|2n|}\}
\end{equation}
so that for each winding number $n$ of $S^2\to S^2$ there are $|2n|$ additional deformation classes. This indicates that there should be additional topological charges which monopole strings which wrap $S^1$ can carry. \\

Indeed, such configurations appear in the BPS spectrum of monopole strings. It is known that in addition to the simple embedding of a monopole string in $\mathbb R^{1+3}\times S^1$ there exists a whole tower of more massive lines \cite{Lee:1997vp, Lee:1998vu, Lambert:2010iw, Tachikawa:2011ch, Kraan:1998kp, Davies:2000nw, Hanany:2001iy}. These can be constructed by realizing that if the fifth component of the gauge field $A_5$ does not depend on $x_5$ then its holonomy acts as an additional adjoint scalar, and the BPS equations become:
\begin{equation}
    B_i=D_i\Phi+D_iA_5
\end{equation}
We may then find a new BPS configuration by constructing $A_5$ related to a standard 't Hooft-Polyakov monopole configuration $\Phi_{mono}$ at asymptotic infinity by a large gauge transformation taking value in the center of $SU(2)$. Explicitly we set:
\begin{equation}
    \Phi = \cos \theta \Phi_{mono} \ \ \ \ \text{and} \ \ \ \ A_5=\sin\theta \Phi_{mono} 
\end{equation}
such that $\sigma^3\cdot \Phi=a\cos\theta$ and $\sigma^3\cdot A_5=a\sin\theta=\frac{2\pi k}{L}$ for real $\theta$ and $k\in \mathbb Z$. This gives the gauge field a holonomy of $(-1)^k\in SU(2)$ around the spatial $S^1$, which we can remove by a large gauge transformation $g(x_5)=\exp(i\pi kx_5\sigma^3/L)$. The resulting states form an infinite tower in $k$ of more massive BPS monopole string solutions. Note that the effect of the gauge transformation is now to yield an asymptotic scalar configuration $\Phi(x_5)=\cos\theta g(x_5)\Phi_{mono}g^{-1}(x_5)$ which has the effect of rotating the configuration $\Phi|_{S^2_{\infty}}:S^2_{\infty}\to SU(2)/U(1)$ by $k$ times around an axis as we circumnavigate $S^1$. For the monopole of minimum magnetic charge $n=1$ and $k=1$ this corresponds to the distinct winding state $(n,l)=(1,1)\in [S^2\times S^1,S^2]$. Thus we see that generalized winding configurations indeed do appear in the BPS spectrum. For nonzero $k$ the above strings also carry UV instanton charge:
\begin{equation}
\frac{1}{8\pi^2}\int_{\mathbb R^3\times S^1}\text{Tr}(G\wedge G)=\frac{1}{8\pi^2}\int \text{Tr}(B_iD_iA_5)=k    
\end{equation}
This is effectively measuring the winding number of $[S^2\times S^1,SU(2)]\cong \mathbb Z$. However since the scalar $\Phi$ is allowed to wind around the full $SU(2)/U(1)$, and since this topological charge contributes to the abelian field strength in the IR, one may ask whether there is some refinement of this symmetry in the Coulomb phase which is capable of distinguishing between the deformation classes of $[S^2\times S^1,SU(2)/U(1)]$. This is the question we will endeavor to answer. \\

\section{Generalized Instanton Symmetry} \label{sec3}

We saw in the previous section that monopole strings which wrap a spatial $S^1$ can have more general winding class by winding asymptotically around $S^2_{\infty}\times S^1$, beyond just their magnetic charge in $[S^2,S^2]$, and moreover that these more general solutions can appear in the BPS spectrum. The UV instanton symmetry generated by $J_I$ can be interpreted as simply evaluating the class in $[S^2\times S^1,SU(2)]\cong \mathbb Z$. However we would like to search for a way to measure the more refined deformation classes in $[S^2\times S^1,SU(2)/U(1)]$ given in \hyperref[deforms]{(2.6)}. \\

For inspiration, we turn to the solitonic charges of the 4d $\mathbb{CP}^1$ $\sigma$-model \cite{Chen:2022cyw, Chen:2023czk, Pace:2023mdo, Freed:2017rlk, Sheckler:2025rlk}, whose degrees of freedom are massless scalar fields $\Sigma : M_4\to \mathbb{CP}^1$ which parametrize $\mathbb{CP}^1\cong S^2$. This theory has two species of solitons: (i) vortices which carry winding number under $\pi_2(S^2)\cong \mathbb Z$ measured by $\Sigma^*\omega_2$, and (ii) particle-like solitons (referred to as \textit{Hopfions}) which carry winding number under $\pi_3(S^2)\cong \mathbb Z$ measured by the Hopf invariant $CS(\Sigma^*\omega_2\wedge \Sigma^*\omega_2)$. \\

Naively one might expect that these two types of solitons should correspond to independent solitonic symmetries, with the vortices charged under a $U(1)^{(1)}$ 1-form symmetry generated by symmetry defect operators $V_{\alpha}(M_2)=\exp(i\alpha\int_{M_2}\Sigma^*\omega_2)$, and the Hopfions charged under a $U(1)^{(0)}$ 0-form symmetry generated by $U_\alpha(M_3)=\exp(i\alpha\int_{M_3}CS(\Sigma^*\omega_2\wedge \Sigma^*\omega_2))$. However it was realized in \cite{Chen:2022cyw} that the presence of the vortices breaks the Hopfion symmetry, and instead the 0-form Hopfion symmetry survives only as a non-invertible $\mathbb{Q/Z}$ symmetry. One manifestation of this is that the vortices can carry Hopfion charge, so that if we surround a vortex of charge $n$ supported on a circle by the defect $U(S^2\times S^1)$ then the resulting Hopf winding number has a $|2n|$-ambiguity and is not well-defined \footnote{It was shown in \cite{Pace:2023mdo} that such a non-invertible symmetry is a generic feature of $\sigma$-models whose target space have multiple homotopy groups which mix nontrivially, in the sense that the corresponding stages of the Postnikov tower are related by a nontrivial fibration. See \cite{Sheckler:2025rlk}  for other examples including gauge theories.} .  \\

The resolution presented in \cite{Chen:2022cyw} is that the symmetry defect operators $U_{\alpha}(M_3)$ should be replaced by non-invertible symmetry defect operators which are capable of measuring the correct $\pi_3$ charge of both the ordinary Hopfions as well as the vortices. These are provided by the minimal abelian TQFTs $\mathcal A^{N,p}$ \cite{Hsin:2018vcg}:
\begin{equation}
    U_{p/N}(M_3)\equiv \mathcal A^{N,p}[M_3,\Sigma^*\omega_2/N]
\end{equation}
for coprime $p,N$, which are coupled to the solitonic 1-form current $\Sigma^*\omega_2/N$ with restricted $\mathbb Z_N$ holonomy. These are the minimal 3d TQFTs with a $\mathbb Z_N$ 1-form symmetry which carry an anomaly labeled by $p$, and the coupling to the bulk 1-form current identifies the 1-form symmetry on the defect worldvolume with a reduction of the 1-form symmetry of the bulk. They form a non-invertible $\mathbb{Q/Z}$ 0-form symmetry \cite{Choi:2022jqy, Cordova:2022ieu, Copetti:2023mcq, Choi:2021kmx, Kaidi:2021xfk}, and are capable of measuring the individual deformation classes of the vortices taking value in $[S^2\times S^1,S^2]$ in \hyperref[deforms]{(2.6)}. \\

This situation bears a clear resemblance to the more general deformation classes of monopole strings wrapping a spatial $S^1$ in the IR of our nonabelian gauge theory. These strings carry magnetic charge in $\pi_2(SU(2)/U(1))$, as well as instanton charge, but we would like to find symmetry defect operators which can measure the full set of topological deformation classes in \hyperref[deforms]{(2.6)}. We can use the above fortuitous capability of the minimal abelian TQFTs $\mathcal A^{N,p}$ in order to construct such codim-1 topological defects. For coprime $p<N$, we define:
\begin{equation}\label{defect}
    \mathcal U_{\frac{p}N}[M_4,\mathcal F/N]= \int Db_2  Dc_1 \exp\Big(i\int_{M_4}\Big[\frac{N}{2\pi}b_2 \wedge dc_1+\frac{Nk}{4\pi}b_2 \wedge b_2+\frac{1}{2\pi}b_2 \wedge \mathcal F\Big]\Big) 
\end{equation}
where $b_2, c_1$ are respectively 2-form and 1-form $U(1)$ gauge fields supported on the defect worldvolume, and $k$ is the multiplicative inverse of $p$  in $\mathbb Z_N$. The worldvolume TQFT of this defect can be considered as a bulk construction of the minimal abelian TQFTs \cite{Hsin:2018vcg, Choi:2022jqy}, in the sense that $\mathcal A^{N,p}$ may be realized on a 3d boundary of $M_4$, and we are coupling to the bulk 't Hooft tensor $\mathcal F/N$ with restricted $\mathbb Z_N$ holonomy. The role of $c_1$ is to act as a Lagrange multiplier which imposes a $\mathbb Z_N$ flux condition on $db_2$, effectively turning $b_2$ into a $\mathbb Z_N$ 2-form connection. Integrating out $b_2$ heuristically yields a 4d SPT phase of the form $\exp\Big(-i\frac{p\pi}N\int_{M_4}\mathcal F\wedge \mathcal F\Big)$, which resembles a fractionally-quantized version of abelian instanton number. More precisely, the bulk worldvolume theory supports the invertible TQFT defined by the Pontryagin square $\frak P(\mathcal F)\in H^4(M_4,\Gamma(\mathbb Z_N))$, where $\Gamma(\mathbb Z_N)$ is the universal quadratic group, so that we can split the defect into bulk and boundary terms:
\begin{equation}\label{split}
    \mathcal U_{\frac{p}N}[M_4,\mathcal F/N]=\exp\Big(\frac{ip\pi}{N}\int_{M_4}\frak P(\mathcal F/N)\Big)\times \mathcal A^{N,p}[\partial M_4, \mathcal F/N]
\end{equation}

We can see how these symmetry defects act on our monopole loops. Suppose that we deform the minimal defect $\mathcal U_{1/N}[\mathbb R^3\times S^1,\mathcal F/N]$ on a spatial slice through a monopole loop carrying some nontrivial winding number in $[S_{\infty}^2\times S^1,S^2]$. We can then evaluate the path integral of the worldvolume theory. As in the instanton calculation, we consider the one-point compactification of spacetime $S^3\times S^1$ with a small chart boundary around infinity of topology $S^2_{\infty}\times S^1$. We can then evaluate the defect on each chart. On the small chart containing the north pole the gauge and Higgs fields are assumed to be trivialized, and so this piece does not contribute. We are then left with the defect $\mathcal U_{1/N}$ evaluated on a space with boundary $S^2_{\infty}\times S^1$. We can then use the fact that the TQFT of \hyperref[defect]{(3.2)} on a space with 3d boundary $\partial M_4=\Sigma_3$ supports a boundary minimal abelian TQFT $\mathcal A_{N,p}[\Sigma_3,\mathcal F/N]$. Thus, since $H^2(\mathbb R^3\times S^1,\mathbb Z)=0$ we can perform a gauge transformation which trivializes $\mathcal F$ in the bulk of the chart, leaving only the boundary contribution $\mathcal A^{N,1}[S_{\infty}^2\times S^1,\mathcal F/N]$. \\

Thus the phase produced by the symmetry defect operator $\mathcal U_{1,N}$ can be reduced to the evaluation of the minimal abelian TQFT $ \mathcal A^{N,1}[S_{\infty}^2\times S^1, \mathcal F/N] $. But the results of \cite{Chen:2022cyw} show that this is precisely the defect which is capable of distinguishing the individual deformation classes in $[S^2\times S^1,S^2]$. In particular, this minimal abelian TQFT can be expressed as the fractional quantum Hall state:
\begin{equation}
    \mathcal A^{N,1}[S_{\infty}^2\times S^1,\mathcal F] = \int \mathcal D a\exp\Big(-i\int_{S_{\infty}^2\times S^1}\frac{N}{4\pi}ada +\frac{1}{2\pi}a\wedge \mathcal F\Big)
\end{equation}
so if the defect surrounds a monopole loop with magnetic charge $n$ then the integral vanishes unless $N|n$, in which case the phase reduces to $\exp\Big(\frac{i}{4\pi^2 N }\int_{S^2\times S^1}CS(\mathcal F\wedge \mathcal F)\Big)$. In a configuration such as that described in \hyperref[sec2]{Section 2} where the only nontrivial winding comes from the scalar then this is precisely the phase $\exp(i\pi l/N)$ where $l \in \mathbb Z_{|2n|}$. \\

By the same argument, if we deform the defect $\mathcal U_{1/N}$ through an instanton of charge $k$ defined by asymptotic wrapping of the scalar around $S^3_{\infty}$ with winding $k$, then the resulting phase is $\exp(i\pi k /N)$, so that these defects are also capable of measuring the instanton charge. Thus we can interpret the symmetry generated by the defects $\mathcal U_{p/N}[M_4,\mathcal F/N]$ as a refinement of instanton symmetry, which is capable of measuring the full class of monopole loop topological charges. \\

We can ask about the invertibility of the symmetry defect operators \hyperref[defect]{(3.2)}. Here there is a critical distinction regarding the boundary conditions for the manifold $M_4$ it is supported on. On a compact manifold with a single global gauge chart, then only the first factor in \hyperref[split]{(3.3)} survives, which is an invertible TQFT, and so the defect operator is invertible. However if we are working on $M_4=\mathbb R^3\times S^1$ with specified boundary conditions along $\partial M_4=S_{\infty}^2\times S^1$ corresponding to the bundle chart boundary, then the fusion of the defect with its inverse will leave a residual contribution supported on the boundary. Indeed, if we take $\mathcal U_{1/N}^{\dagger}[M_4]\equiv\mathcal U_{-1/N}[M_4]$ then it is a straightfoward calculation \footnote{This can be achieved by enforcing the worldvolume gauge field equations of motion after a field redefinition, as in \cite{Choi:2022jqy, Kaidi:2021xfk, Choi:2021kmx}} to verify that:
\begin{equation}
    \mathcal U_{1/N}[M_4,\mathcal F]\times\mathcal U_{1/N}^{\dagger}[M_4,\mathcal F]=\mathcal C[\partial M_4,\mathcal F]
\end{equation}
where $\mathcal C[\Sigma_3,\mathcal F]$ is a condensation defect \cite{Choi:2021kmx, Choi:2022jqy, Kaidi:2021xfk} coupled to the bulk field strength $\mathcal F$. Thus we see that on a manifold with specified boundary conditions, the generalized instanton symmetry defect operators are in fact non-invertible, but satisfy an augmentation of the fusion rules of familiar categorical $\mathbb{Q/Z}$ symmetry.

\section{Origin from the 6d $\mathcal N=(2,0)$ Theory}\label{sec4}

A common theme in the study of symmetries is that, very often, symmetry structures in a quantum field theory might have some manifest geometric or topological interpretation when approached from an alternative perspective of string theory, holography, or some proposed UV completion. This is especially true in the context of supersymmetric gauge theories, which sometimes admit constructions from dimensional reduction of higher-dimensional theories, or geometric engineering. It is thus natural to wonder whether the generalized instanton symmetry discussed in the previous section has such an origin. \\

A natural setting where this is the case, as we shall show, is the 5d $\mathcal N=2$ theory with gauge group $SU(2)$, which has maximal supersymmetry. There are several reasons why this is a convenient example. First, while this theory is itself non-renormalizable and so is not perturbatively well-defined, it can be thought of as a low-energy effective theory arising from the 6d $\mathcal N=(2,0)$ theory of type $\frak a_1$ under compactification on a circle $S^1$. There is a long list of phenomena in supersymmetric gauge theories which can be explained as arising from the data of the 6d $(2,0)$ theories under dimensional reduction.  Another appealing feature is that the maximal supersymmetry protects the Coulomb branch of the 5d theory from dramatic quantum corrections, making the uplift to 6d tractable. In this section we will study the relationship between the 5d and 6d theories and see that our generalized instanton symmetry of interest indeed has a natural origin. 

\subsection{Symmetries of the 6d $\mathcal N=(2,0)$ Theory}

First let us recall some features of the symmetries of the 6d $\mathcal N=(2,0)$ theories of type $\frak{a}_{N-1}$. For a more thorough review see e.g. \cite{Moore_FK, Heckman:2018jxk}. These theories have various constructions, either from geometric engineering in Type IIB supeerstrings or as the worldvolume theory on a stack of $N$ M5-branes in M-theory \cite{Witten:1995zh, Seiberg:1996qx, Strominger:1995ac,Witten:1996hc}. These are relative quantum field theories \cite{Freed:2012bs}, meaning their partition functions are not themselves well defined, but must be defined as the 6d boundary theory of a non-invertible 7d topological field theory known as the \textit{anomaly theory} \cite{Witten:1996hc, Witten:1998wy, Gukov:2020btk}. This means that the theory is actually defined using a ``partition vector" \cite{Tachikawa:2013hya} in the space of conformal blocks, requiring some choice of polarization. The local form of the anomaly theory has been well-studied \cite{Harvey:1998bx, Intriligator:2000eq, Maxfield:2012aw, Ohmori:2014kda}, while the full global form of the anomaly is discussed in \cite{Monnier:2017klz, Monnier:2014txa, Sheckler:2025lfv}. \\

The non-invertible piece of the 7d anomaly theory is a shifted Wu-Chern-Simons theory \cite{Monnier:2016jlo, Monnier:2017klz}, which is the higher-dimensional analogue of a spin Chern-Simons theory. This can be constructed as a 7d topological field theory in terms of a 3-form shifted differential cocycle $\check C_3$. When this differential cocycle is topologically trivial then the action can be expressed in the familiar local form \cite{Gukov:2020btk}:
\begin{equation} \label{anom}
\frac{\Omega_{ij}}{4\pi}\int_{M_7} C_3^i\wedge dC_3^j
\end{equation}
where $\Omega_{ij}$ is the Cartan matrix of $\frak a_{N-1}$. The 6d (2,0) theory arises as a 6d boundary theory for this TQFT, with the Hilbert space composed of states prepared by the 7d bulk along with the physical 6d boundary conditions. The description of such a state requires the specification of a maximal isotropic subgroup of $H^3(M_6,\mathbb Z)$ corresponding to a choice of basis. \\

The moduli space of vacua of the 6d (2,0) theory of type $\frak a_{N-1}$ is easier to describe. A generic point on the tensor branch is defined by giving vevs to the 5 scalars parameterizing the separation of the M5-branes, and the theory here is described by $N-1$ free tensor multiplets. These each contain a 2-form $U(1)$ gauge field $b_2^i$ with self-dual field strength $h_3^i=*h_3^i$. Also present are self-dual strings with nonzero tension which carry both electric and magnetic charge under $h_3$, so that the global 2-form symmetry is $\mathbb Z_N^{(2)}$  \cite{DelZotto:2015isa} (for related discussions on higher form symmetries in 6d SCFTs see e.g. \cite{Bhardwaj:2020phs, Apruzzi:2022dlm, Cordova:2020tij, Apruzzi:2021vcu, Apruzzi:2020zot, Apruzzi:2024cty, Bonetti:2024etn}) . The self-dual field strengths have a natural relationship to the 7d anomaly TQFT as being the required boundary values of the bulk differential cocycles $\check C^i_3|_{\partial M_7}=h^i_3=db^i_2 $.

\subsection{Dimensional Reduction to 5d}

The dimensional reduction of the 6d $\mathcal N=(2,0)$ theory of type $\frak a_{N-1}$ on a circle $S^1$ yields 5d $\mathcal N=2$ super-Yang-Mills with gauge group $G=SU(N)$. This theory is non-renormalizable and so should be treated as a low-energy effective theory with this particular UV completion. The gauge coupling is related to the compactification radius $g^2 = 4\pi^2 R$. There has been extensive work on understanding whether this 5d theory can be considered as complete or whether additional non-perturbative information is needed to recover the UV 6d theory \cite{Douglas:2010iu, Lambert:2010iw, Kim:2011mv, Cordova:2015vwa, Duan:2021ges, Lambert:2010wm}. \\

For the 5d theory with Lie algebra $\frak g$ then the scalar potential in the action restricts the adjoint scalars to the Cartan $\frak {t_g} \subset \frak g$ in the IR. Thus the Coulomb branch is parametrized by $\braket{\phi^I}\in \mathbb R^5\otimes \frak {t_g}/\mathcal W_{\frak g}$ where $\mathbb R^5$ rotates under $SO(5)_R$ and $\mathcal W_{\frak g}$ is the Weyl group. Since these scalars are inherited from the tensor multiplets then the Coulomb branch corresponds to the 6d tensor branch under dimensional reduction on $S^1$. \\

While there can be higher-derivative corrections to the leading-order two-derivative effective action, these are under control. On the other hand the maximal supersymmetry protects the leading effective action from quantum corrections \cite{Cordova:2015vwa, Paban:1998ea, Paban:1998qy, Sethi:1999qv, Movshev:2009ba}, and so the geometry of the Coulomb branch is completely described by the classical theory. This allows us to extrapolate up to energies above the KK scale $\frac 1R$ and provide an explicit matching between degrees of freedom on the Coulomb branch with those on the 6d tensor branch. In particular, we can identify the field strengths $f_i$ of the abelian gauge fields on the Coulomb branch as assembling into the self-dual field strengths as \cite{Cordova:2015vwa}:
\begin{equation} \label{FS}
    h_3^i\sim \frac{1}{2\pi R}(f_i\wedge dx_5+*_5f_i)
\end{equation}
where $x_5$ is the coordinate of the compactified circle and $*_5$ is the Hodge star in the other five dimensions. \\

We can also consider the $\frac 12$-BPS objects on the Coulomb branch, which are the electrically-charged W-bosons labeled by roots in $\Delta_{\frak g}$ and the magnetically-charged monopole strings labeled by coroots. These are expected to arise from the self-dual strings on the 6d tensor branch under dimensional reduction, which have the appropriate tension. W-bosons arise from self-dual strings which wrap the compactified circle, while monopole strings arise from unwrapped strings. This gives them the appropriate electric and magnetic charges under \hyperref[FS]{(4.2)}, and realizes S-duality \cite{Tachikawa:2011ch}. The tower of higher-mass monopole strings discussed in \hyperref[sec2]{Section 2} which carry instanton charge can be seen as deriving from unwrapped self-dual strings with some nonzero Kaluza-Klein momentum. \\

Since the 6d theory is only defined as a theory relative to the 7d anomaly theory \hyperref[anom]{(4.1)}, what becomes of this bulk theory under compactification? Consider the case of $G=SU(2)$ which has just a single gauge field with 7d TFT $\frac{2}{4\pi}\int_{M_7}C_3\wedge dC_3$. Then under dimensional reduction the $C$-field decomposes into two distinct gauge fields $C_3=B_3+\frac{dx_5}{2\pi}\wedge B_2$ so that the 5d theory now has a 6d bulk \cite{Gukov:2020btk}:
\begin{equation}\label{su2anom}
    \frac{2}{2\pi}\int B_3\wedge dB_2
\end{equation}
This is an invertible BF-theory describing a $\mathbb Z_2$ 1-form symmetry, so we see that instead of the 5d theory being relative it is in fact an absolute theory. More generally, the dimensional reduction of \hyperref[anom]{(4.1)} for the theory of type $\frak a_{N-1}$ yields a BF-type theory describing the center $\mathcal Z(SU(N))=\mathbb Z_N$ 1-form symmetry of the 5d SYM \cite{DelZotto:2015isa}. We discuss other gauge groups in \hyperref[sec5]{Section 5}. \\

\subsection{Instanton Symmetry and Defects}

In addition to the symmetries inherited from the 6d $\mathcal N=(2,0)$ theory, there is a new global symmetry in the 5d theory given by the Kaluza-Klein symmetry arising from the isometry of the compactified circle. The KK modes are identified with the instantons of the 5d theory, and so the $U(1)$ KK symmetry is identified with the $U(1)_I$ instanton symmetry. This can be seen explicitly, for example, from the reduction of the supercurrents in the 6d superconformal algebra \cite{Lambert:2010iw}, or by comparison of the KK and instanton energy scales and zero-modes $\cite{Kim:2011mv, Tachikawa:2011ch}$. \\

We would like to understand how the generalized refinement of instanton symmetry on the Coulomb branch which we found in the previous section fits into this interpretation as KK symmetry. For this, it is useful to once again study the manifestation of the symmetries in the 6d bulk topological theory, with the 5d theory defined as the physical boundary. Thus far we have referred to this theory as the anomaly theory, but there exists a modern adaptation of this known as the Symmetry Topological Field Theory (SymTFT) \cite{Freed:2022qnc, Bhardwaj:2023ayw, Kapustin:2014gua, Ji:2019jhk, Apruzzi:2021nmk, Gaiotto:2020iye}. This involves truncating the bulk by adding an additional boundary with a chosen topological boundary condition, forming a slab in which the bulk TQFT lives. It is proposed that the SymTFT along with the data of the boundary conditions should contain all information about the symmetries of the physical boundary theory, including all charges and anomalies. Beyond just the anomaly theory, the SymTFTs for 6d $\mathcal N=(2,0)$ theories and their dimensional reductions have been used in some cases to study generalized symmetries using the holographic formulation \cite{Bonetti:2024etn, Apruzzi:2024cty, Lawrie:2023tdz, Bashmakov:2022uek}. \\

If we want to study the details of the KK instanton symmetry and its relationship to the other symmetries of the theory, it is therefore useful to understand its contribution to the full SymTFT. In order to do this, we must include a background $U(1)$ gauge field $A_1$ for this symmetry in the bulk 7d TFT as we perform the dimensional reduction down to 6d. However recall that this symmetry in 6d is interpreted as the isometry of the $S^1$. Following \cite{Gukov:2020btk}, our decomposition of the 7d $C_3$ field must be modified to 
\begin{equation}
    C_3=B_3+\frac{dx_5-A_1}{2\pi}\wedge B_2
\end{equation}
Then gauge invariance of the field strength $dC_3$ requires that the dimensionally-reduced gauge fields transform as a 3-group \cite{Cordova:2018cvg, Benini:2018reh}: 
\begin{equation}
    A_1\to A_1+d\lambda \ \ \  \ , \ \ \ \  B_2\to B_2+d\Lambda_1 \  \ \ \ , \  \ \ \ B_3\to B_2+d\Lambda_2+dA_1\wedge \Lambda_1
\end{equation}
The associated gauge invariant field strength of the dimensionally reduced field is therefore the combination $dB_3-\frac{1}{2\pi}dA\wedge B_2$. Thus after including the $A_1$ gauge field, we see that in the case of $G=SU(2)$ the 6d anomaly theory \hyperref[su2anom]{(4.3)} obtained from compactification receives an additional term:
\begin{equation}
    \frac{2}{2\pi}\int_{M_6} \Big(B_3\wedge dB_2-\frac{A_1}{2\pi}\wedge B_2\wedge dB_2\Big)
\end{equation}
The new term describes a mixed anomaly between the center $\mathbb Z_2$ 1-form symmetry and the $U(1)_I$ instanton symmetry \cite{BenettiGenolini:2020doj, Sheckler:2025rlk, Apruzzi:2022dlm}. In order to fully incorporate the $U(1)_I$ symmetry into the SymTFT we need to include a term corresponding to the SymTFT for a $U(1)$ symmetry \cite{Brennan:2024fgj, Antinucci:2024zjp}, so that the full TQFT action is:
\begin{equation}\label{su2symtft}
    \int_{M_6}\frac{2}{2\pi}B_3\wedge dB_2-\frac{2}{2\pi} B_2\wedge dB_2\wedge\frac{A_1}{2\pi}+\frac{1}{2\pi}H_4\wedge dA_1
\end{equation}
where $H_4$ is a $\mathbb R$-valued gauge field. This is the same SymTFT which was recently studied in \cite{Bertolini:2025wyj}, where the topological boundary conditions were classified and it was used to study the instanton symmetry for different global forms of $SU(N)/\mathbb Z_k$ gauge theory in 5d. \\

There is one more feature of the bulk theory which is also crucial to include. Recall that the tensor branch of the 6d (2,0) theory possesses self-dual strings which carry both electric and magnetic charge under $db_2=h_3$. This means that in the presence of such defects the 2-form field $b_2$ should be modeled as a \textit{non-closed} differential cochain $\check b_2\in \check C^3(M_6)$ which is sourced by these defects. But recall that the boundary condition for the 7d anomaly theory was that the bulk differential cocycle $\check C_3$ appearing in the Wu-Chern-Simons theory had to coincide with the field strength on the boundary $\check C_3|_{M_6}=d\check b_2$. This means that in the presence of defects $\check C_3$ is necessarily a topologically non-trivial differential cocycle, and thus must have nontrivial holonomy around cycles which reduce to cycles linking the defect on the boundary \cite{Monnier:2017klz}. In particular, this means that the defects representing the self-dual strings must necessarily extend into the 7-dim bulk in order to provide cycles for such holonomy. This is consistent with the M-theory realization of the 6d (2,0) theory, where the self-dual strings on the tensor branch are realized as the ends of M2-branes stretching between adjacent M5-branes. \\

Now we can study the $U(1)_I$ instanton symmetry on the Coulomb branch of the 5d theory from the bulk perspective. We are interested in determining the topological defect operators which generate this symmetry. The naive topological operators corresponding to just the $U(1)_I$ SymTFT piece of the topological action are the Wilson and 't Hooft operators: $W_n(\gamma)=\exp(in\oint_{\gamma}A_1)$ for $n \in \mathbb Z$ and $V_{\alpha}(\Gamma)=\exp(i\alpha\oint_{\Gamma}H_4)$ for $\alpha \in \mathbb{R/Z}$. Since $U(1)_I$ is an ungauged global symmetry on the boundary then the former are given Dirichlet boundary conditions, so that we expect $V_{\alpha}(\Gamma)$ to provide the symmetry defect operators for $U(1)_I$ when taken to the boundary. For this to be the case these operators must be topological. \\

However we notice that the equation of motion for $A_1$ picks up a contribution from the anomaly term:
\begin{equation}
    dH_4=\frac{2}{2\pi}B_2\wedge dB_2
\end{equation}
This means that if we deform $V_{\alpha}(\Gamma)$ through a 5d bulk $\Sigma_5$ then the contribution to the action is a phase:
\begin{equation}\label{su2var}
    \Delta=\exp\Big(i\frac{2\alpha}{2\pi}\int_{\Sigma_5}B_2\wedge dB_2\Big)
\end{equation}
There is an obvious problem with this phase: it is not manifestly gauge invariant. Indeed, under a gauge transformation $B_2\to B_2+\Lambda_2$ then this phase has a variation:
\begin{equation} \label{var}
    \delta_{\Lambda_2}\Delta=\exp\Big(i\frac{2\alpha}{2\pi}\int_{\Sigma_5}\Lambda_2\wedge dB_2\Big)
\end{equation}
Moreover, we know that the defects corresponding to monopole strings must be extended out into the bulk along some submanifold $N_3$, so we can consider situations where the deformation through $\Sigma_5$ intersects this defect. Since we have the boundary condition that $B_2$ reduces to the abelian gauge field strength on the boundary $B_2|_{M_5}=\mathcal F$ then it is sourced by these defects:
\begin{equation}
 dB_2=n\delta^{(3)}(N_3)   
\end{equation}
where $\delta^{(3)}(N_3)$ is the 3-form Poincare dual to $N_3$. Therefore, in general, the above gauge-variation \hyperref[var]{(4.10)} need not vanish. For example, if the extended defect wraps a 2-cycle in the bulk then $\Lambda_2$ can be chosen to be a large gauge transformation around this cycle. This could be the case, for example, if in addition to wrapping a spatial $S^1$ we also consider our monopole string in a theory with finite temperature. If we allow for such fully general topologies and configurations in the analysis of our SymTFT, then the phase $\Delta$ and its variation are only \textit{guaranteed} to vanish if $\alpha = k/2n$ for some $k \in \mathbb Z$. \\

Thus we find that, in the presence of a monopole string of charge $n$, then the symmetry defect operators $V_{\alpha}(\Gamma)$ of the SymTFT are only \textit{always} defined for $\alpha \in \mathbb {Z}_{|2n|}\subset\mathbb{Q/Z}$. Moreover, the defect must be ``smart" enough to detect the charge $n$. These are precisely the capabilities which characterize the defects \hyperref[defect]{(3.2)} which we found in \hyperref[sec3]{Section 3}. Thus we see that the dimensional reduction for the 6d (2,0) theory of type $\frak a_1$ which correctly takes the KK symmetry and self-dual strings into account, as well as the correct boundary conditions, gives rise to a SymTFT which naturally supports our generalized instanton symmetry defect operators. \\

Of course, for the specific geometries we are considering, of a monopole string wrapping only an $S^1$ and symmetry defect operator passing through it transversely, then there is no issue of gauge invariance for $V_{\alpha}(\Gamma)$ for $\alpha \in \mathbb{R/Z}$. This reflects the fact that the full $U(1)_I$ symmetry is still a valid symmetry in the Coulomb phase. Nevertheless, we see that our generalized refinement of this symmetry is in fact the more general version, and is naturally predicted by the theory from the SymTFT perspective. These are the defects which are capable of measuring the full collection of winding configurations for the monopole strings.  \\

In this argument we have focused on the case of maximal supersymmetry, because in this case we were able to explicitly derive the anomaly in the SymTFT from dimensional reduction of the 6d (2,0) theory, and because the matching of the Coulomb and tensor branches gives an explicit argument for why the defects corresponding to self-dual strings must necessarily be extended into the bulk. However we expect that this argument should apply generically for any Coulomb phase of a 5d $SU(2)$ gauge theory, since there exist general arguments for the existence of such a mixed anomaly between the center 1-form symmetry and the instanton symmetry \cite{BenettiGenolini:2020doj}. Indeed, the SymTFT we derived here was taken in \cite{Bertolini:2025wyj} as the correct theory by assumption. An alternative argument for the existence and form of the mixed anomaly is in \cite{Sheckler:2025rlk}, since the center 1-form symmetry is an electric symmetry and the instanton symmetry is a magnetic symmetry corresponding to the lowest homotopy group of $ SU(2)$. Moreover, an advantage of taking the perspective of the 6d (2,0) theory is that it will facilitate a path towards generalization to other gauge groups, which we undertake next. 

\section{Generalization to Other Gauge Groups} \label{sec5}

Thus far we have restricted our discussion to the case of $G=SU(2)$, so that the SSB of the gauge symmetry is down to only a single $U(1)$. We would of course like to understand the case of general semi-simple compact Lie group $G$. Adjoint scalars can then spontaneously break the gauge symmetry down to $G\to T_G$ where $T_G$ is generated by the Cartan. The ordinary monopole solutions come from topologically-nontrivial winding in $\pi_2(G/T_G)$, and the Dirac quantization condition enforces that the possible magnetic charges belong to the co-root lattice $\vec n \in \Lambda_{co-root}^G$. In 5d we can consider monopole strings which wrap a spatial cycle, and thus can carry generalized winding configurations classified by:
\begin{equation}\label{genhom}
    [S^2\times S^1,G/T_G]
\end{equation}
We can therefore ask if there exists a refinement of the instanton symmetry, analogous to what we found for $G=SU(2)$ and generated by the defects \hyperref[defect]{(3.2)}, which is capable of measuring this more general topological charge. A direct generalization would require a full analysis of the homotopy classes in \hyperref[genhom]{(5.1)}, which we do not undertake in this work. However the results of the previous section give us another avenue for investigation. The fact that we were able to detect the existence of these generalized instanton symmetry defects by dimensionally reducing the anomaly TFT of the 6d $(2,0)$ theory of type $\frak a_1$ means that perhaps we can perform the same procedure for any simply-laced ADE Lie algebra $\frak g$. This is the approach we shall pursue.\\

We can then perform the same dimensional reduction on $S^1$ as before, decomposing the fields as $C_3^i=B_3^i+\frac{1}{2\pi}(dx_5-A_1)\wedge B_2^i$, where we have included a possible background for the KK symmetry on $S^1$. After including the SymTFT contribution for the instanton symmetry, this describes the relevant contributions to the SymTFT given by:
\begin{equation}
     i\int_{M_6}\sum_{i,j}^{r(\frak g)}\Omega_{ij}\Big[\frac{1}{2\pi}B_3^i\wedge dB_2^j-\frac{1}{2\pi} B_2^i\wedge dB_2^j\wedge\frac{A_1}{2\pi}\Big]+\frac{1}{2\pi}H_4\wedge dA_1
\end{equation}
which is a simple generalization of \hyperref[su2symtft]{(4.7)}. Once again we see that the anomaly contributes to the topological condition for the instanton symmetry defect operators via the equation of motion:
\begin{equation}\label{genanom}
    dH_4=\frac{1}{2\pi}\Omega_{ij}B_2^i\wedge dB_2^j
\end{equation}
By the same argument as before, when the naive instanton symmetry defect operators $\exp(i\alpha\int H_4)$ are deformed through a monopole string defect extended out into the bulk, the resulting phase will not be gauge invariant in a generic topology. We again take this to mean that the most general form of these defects must therefore be restricted to certain discrete values of $\alpha$ which depend intelligently on the monopole string charge. \\

We would like to understand what these non-invertible symmetry defect operators might look like. For this, we take lessons from the $SU(2)$ case and will see that there is a natural generalization. In this case, the defects $\mathcal U_{p/N}(M_4,B)$ in \hyperref[defect]{(3.2)} were defined by hosting a 4d TQFT on the worldvolume given by the Pontryagin square: 
\begin{equation}\label{su2ps}
    \exp\Big(\frac{ip\pi}{N}\int_{M_4}\frak P(B)\Big)\times \mathcal A^{N,p}[\partial M_4, B]
\end{equation}
with a 3d minimal abelian abelian TQFT $\mathcal A^{N,p}[\partial M_4, B]$ living on a possible boundary. The bulk field strength $B$ was taken to be a reduced $\mathbb Z_N$ gauge field $f$, and the ability of these defects to measure generalized homotopy classes of monopole strings resulted from the capability of $\mathcal A^{N,1}[B]$ to distinguish the classes in $[S^2\times S^1,SU(2)/U(1)]$ when $N$ was taken to coincide with magnetic charge $n$. \\

The Pontryagin square is a cohomology operation $\frak  P :H^2(\Gamma_4,\mathbb Z_N)\to H^4(\Gamma_4,\Gamma(\mathbb Z_N))$ where $\Gamma(\mathbb Z_N)$ is the universal quadratic group given by $\mathbb Z_N$ if $N$ is odd and $\mathbb Z_{2N}$ if $N$ is even. In fact there exists a more general formulation of such a TQFT built from the Pontryagin square (see e.g. \cite{Benini:2018reh, Aharony:2013hda, Kapustin:2013qsa} for a review). If $\mathcal A$ is any abelian group, and $\Gamma(\mathcal A)$ is the universal quadratic group, then the Pontryagin square is a cohomology operation $\frak P : H^2(X,\mathcal A)\to H^4(X,\Gamma(\mathcal A))$. Then given any quadratic form $h : \mathcal A\to \mathbb{R/Z}$ there exists a unique lift $\tilde h : \Gamma(\mathcal A)\to \mathbb{R/Z}$, which allows us to construct the action: 
\begin{equation}
    \exp\Big(2\pi i\int_{M_4}\tilde h(\frak P(B))\Big)
\end{equation}
If we then consider the TQFT \hyperref[su2ps]{(5.4)} for the $SU(2)$ case as defining the $\mathbb Z_N$ reduction of the topological term $2 B_2\wedge B_2$ appearing in the anomalous phase \hyperref[su2var]{(4.9)} for $\mathbb Z_N$-reduced gauge fields $B_2$, then the natural generalization to the case of general $\frak g$ is to consider the $\mathbb Z_{N_i}$ reduction of $B_2^i$ for all $r(\frak g)$ gauge fields, and take the reduction of the TQFT appearing in \hyperref[genanom]{(5.3)}, using the quadratic form $\Omega^{ij}$. In particular, we take $\mathcal A=\prod_i^{r(\frak g)}\mathbb Z_{N_i}$ and use the quadratic form:
\begin{equation}
    h[\{a_i\}]\equiv \sum_{i,j}\frac{1}{2N_{ij}}a_i\Omega_{ij}a_j \ \  \text{for } \  \vec a\in \mathcal A \ \  \text{and} \  \ N_{ij}\equiv \gcd(N_i,N_j)
\end{equation}
The resulting TQFT can then be expressed as \cite{Hsin:2018vcg}:
\begin{equation}
    \exp\Big(2\pi i\int_{M_4}\Big[\sum_i\frac{\Omega_{ii}}{2N_i}\frak P(B^i)+\sum_{i<j}\frac{\Omega_{ij}}{N_{ij}}B^i\cup B^j\Big]\Big)
\end{equation}
We can express this in terms of bulk $U(1)$ gauge fields $c_1^i, b_2^i$, which allows us to define our generalized symmetry defect operators:
\begin{equation}\label{genop}
    \mathcal U_{N_1,\dots, N_{r(\frak g)}}[M_4,\{B_2^i\}]\equiv \int \prod_i^{r(\frak g)}Db_2^iDc_1^i\exp\Big(i\int_{M_4} \frac{1}{2\pi}\sum_iN_idc_1^ib_2^i+\frac{2}{2\pi}\sum_{i,j}\frac{N_iN_j}{N_{ij}}\Omega_{ij}[b_2^ib_2^j+b_2^iB_2^j]\Big)
\end{equation}
Applying the equations of motion reproduces the above worldvolume TQFT. This theory has a combined 1-form and 2-form gauge symmetry: 
\begin{equation}
    b_2^i\to b_2^i-d\Lambda_1^i \ \ \ \ , \ \ \ \ \ c_1^i\to c_1^i+\sum_j\frac{N_j}{N_{ij}}\Omega_{ij}\Lambda_1^j
\end{equation}
For simplicity we do not consider more general versions of the theory with more general anomalies. The gauge invariant operators are surface operators:
\begin{equation}
    V_{\vec a}(\Sigma)=\exp\Big(i\int_{\Sigma}\sum_{i,j}a_i\frac{\Omega_{ij}N_j}{N_{ij}}b_2^j+i\oint_{\partial \Sigma}\sum_ia_ic_1^i\Big)
\end{equation}
We are interested in the 3d boundary theory on $\partial \Gamma_4=M_3$ when we impose Dirichlet boundary conditions $b_2^i|_{M_3}=0$, which explicitly breaks the gauge symmetry. This means that the line operators $W_{\vec a}(\gamma)=\exp(i\oint_{\gamma}\sum_ia_ic_1^i)$ become gauge-invariant topological operators on the boundary. Their braiding is defined by the symmetric bilinear form $\mathcal A\times \mathcal A\to \mathbb{R/Z}$ associated to the quadratic form $h[\vec a]$, given by:
\begin{equation}
    \exp(i(h(aa')-h(a)-h(a'))=\exp\Big(2\pi i\int\sum_{i,j}\frac{1}{N_{ij}}a_i\Omega_{ij}a_j\Big)
\end{equation}
This bilinear form defines a map $M:\mathcal A\to\widehat{\mathcal A}=\text{Hom}(\mathcal A,U(1))$ between $\mathcal A$ and its Pontryagin dual. When this map has trivial kernel then every line braids nontrivially, and so the boundary theory preserves the 1-form symmetry $\mathcal A$, and thus defines a modular TQFT. The source field $B^i$ couples to these lines on the boundary since $B^i$ couples to $\sum_j\frac{\Omega_{ij}N_j}{N_{ij}}b_2^j$ in the bulk, and the surfaces $\exp\Big(i\int \sum_j\frac{\Omega_{ij}N_j}{N_{ij}}b_2^j\Big)$ end as $\exp(i\oint c^i)$ on the boundary. \\

We can therefore define the boundary TQFT $\mathcal A_{N_1,\dots, N_R}[M_3,\{B_2^i\}]$ generated by these lines as a generalization of the minimal abelian TQFTs $\mathcal A^{N,1}$, which they reduce back to in the case of $SU(2)$. In this case, the capability of the defects \hyperref[defect]{(3.2)} to distinguish the general winding classes of the monopole lines came from the capability of $\mathcal A^{N,1}$ to measure the winding classes in $[S^2\times S^1,SU(2)/U(1)]$ when $N$ was set equal to the magnetic charge $n$. We therefore find it tempting to propose that the TQFTs $\mathcal A_{N_1,\dots, N_R}[M_3,\{B_2^i\}]$ also provide homotopy invariants, capable of distinguishing the homotopy classes in $[S^2\times S^1,G/T_G]$. A thorough confirmation of this would require a detailed study of these homotopy classes, which we leave to future work. The ``intelligence" of these defects would be reflected in the fact that they only measure the correct charge when the parameters $N_i$ are set equal to the magnetic charges $\vec n\in \Lambda_{\frak g}$ of the monopole string. Thus it is sufficient to only consider these TQFTs for parameters in the root lattice $\vec N\in \Lambda_{\frak g}$. 

\section{Discussion} \label{sec6}

In this work we have pointed out that there exists a generalization of instanton symmetry which is capable of detecting a larger set of winding classes for magnetic monopole strings in 5d. The symmetry defect operators for this symmetry are non-invertible when placed on a manifold with boundary, but otherwise are invertible and support a 4d TQFT coupled to the abelian field strength in the IR after SSB of the gauge symmetry. We have paid particular attention to the case of maximal supersymmetry, since in this case the origin of these defects can be connected to the SymTFT obtained from dimensional reduction of the anomaly TFT for the 6d $\mathcal N = (2,0)$ SCFT. While the construction of this symmetry in the case of $G=SU(2)$ is straightforward, there are several remaining questions and directions which give prospects for future work. \\

The dynamical role of this symmetry and the associated generalized charges for the monopole strings remains open. We have seen in \hyperref[sec2]{Section 2} that monopole lines which carry generalized instanton number can appear in the BPS spectrum, but it would be fruitful to understand how the full spectrum of lines is organized in terms of these topological charges. One interesting avenue for investigation exists in the maximally-supersymmetric case of the 5d $\mathcal N=2$ theory, where the ordinary instanton symmetry can be directly interpreted as the KK symmetry arising from the dimensional reduction of the 6d $(2,0)$ theory. In this case, there is an intricate and mysterious relationship between the instanton physics and the details of this UV completion \cite{Douglas:2010iu, Lambert:2010iw, Kim:2011mv, Cordova:2015vwa, Duan:2021ges, Lambert:2010wm}. Studies of instanton zero modes and string theory realizations of instantons as threshold bound states of D0-D4 branes have led to the hypothesis that an instanton of 5d $SU(N)$ gauge theory is actually composed of $N$ partons which each carry fractional $1/N$ instanton number \cite{Bolognesi:2011nh, Collie:2009iz, Kim:2011mv, Lee:1997vp, Boyarsky:2002ck}. This has interesting implications to the curious $N^3$ scaling of the anomaly coefficient for the $A_N$ 6d (2,0) theories. However an instantonic gauge field configuration as considered in \cite{Bolognesi:2011nh} can properly only have fractional instanton number if the global form of the gauge group is some $SU(N)/\mathbb {Z}_k$ obtained by gauging some subgroup of the $\mathbb Z_k\subset \mathbb Z_N $ 1-form center symmetry. On the other hand, we saw in this work that our generalized instanton symmetry behaves heuristically like a fractional instanton symmetry, but the generalized instanton number is carried by the monopole strings with finite KK momentum. It would be fascinating to better understand the role such charges can play in the limit of the fully-symmetric theory, and how this contributes to the 6d UV completion. \\

Another direction for exploration is a more thorough description of the 6d SymTFT for the 5d theory obtained from the 6d (2,0) theory. In particular, we have used only the non-invertible piece \hyperref[anom]{(4.1)} of the full 7d topological anomaly TFT described in \cite{Monnier:2017klz}, neglecting both the local and global invertible contributions. Moreover, the action \hyperref[anom]{(4.1)} defines the correct TQFT only when the bulk gauge field is topologically trivial, but as we have stressed, the existence of the dynamical self-dual strings which must be extended as defects into the 7d bulk make this a topologically-nontrivial differential cocycle. A proper analysis should then use the formulation of this non-invertible piece as a shifted Wu-Chern-Simons theory \cite{Monnier:2016jlo} associated to the lattice $\Lambda_{\frak g}$. It would also be useful to develop a proper holographic understanding for how exactly the $U(1)_I$ instanton symmetry SymTFT arises in a way which is compatible with the usual derivation of the anomaly TFT. The fact that in 6d this symmetry is a spacetime isometry symmetry of the compact $S^1$ makes such a description challenging, although there has been recent work on developing SymTFT descriptions for spacetime symmetries \cite{Apruzzi:2025hvs}. \\

Finally, while we have focused primarily on the case of $G=SU(2)$, it would be interesting to further study the generalization to arbitrary ADE gauge group. This could be achieved by developing a direct generalization of the defects \hyperref[defect]{(3.2)} which detect individual deformation classes in $[S^2\times S^1,G/T_G]$, or by a more detailed study of the proposal in \hyperref[sec5]{Section 5}. The latter approach requires both a more thorough description of the symmetry defect operators \hyperref[genop]{(5.8)} and their boundary TQFTs, as well as a better understanding of their ability to act as homotopy-invariants. In either case, a description of the full set of homotopy classes in $[S^2\times S^1,G/T_G]$ would be instrumental.

\acknowledgments

AS is supported by Simons Foundation award 568420 (Simons Investigator) and award 888994 (The Simons Collaboration on Global Categorical Symmetries). It is my pleasure to thank Ken Intriligator, Daniel Brennan, Clay Cordova, and Zipei Zhang for helpful discussions.

\bibliographystyle{JHEP}
\bibliography{biblio.bib}

\providecommand{\href}[2]{#2}\begingroup\raggedright\begin{thebibliography}{10}

\bibitem{Gaiotto:2014kfa}
D.~Gaiotto, A.~Kapustin, N.~Seiberg and B.~Willett, \emph{{Generalized Global Symmetries}}, \href{https://doi.org/10.1007/JHEP02(2015)172}{\emph{JHEP} {\bfseries 02} (2015) 172} [\href{https://arxiv.org/abs/1412.5148}{{\ttfamily 1412.5148}}].

\bibitem{Cordova:2022ruw}
C.~Cordova, T.T.~Dumitrescu, K.~Intriligator and S.-H.~Shao, \emph{{Snowmass White Paper: Generalized Symmetries in Quantum Field Theory and Beyond}},  in \emph{{Snowmass 2021}}, 5, 2022 [\href{https://arxiv.org/abs/2205.09545}{{\ttfamily 2205.09545}}].

\bibitem{McGreevy:2022oyu}
J.~McGreevy, \emph{{Generalized Symmetries in Condensed Matter}}, \href{https://doi.org/10.1146/annurev-conmatphys-040721-021029}{\emph{Ann. Rev. Condensed Matter Phys.} {\bfseries 14} (2023) 57} [\href{https://arxiv.org/abs/2204.03045}{{\ttfamily 2204.03045}}].

\bibitem{Brennan:2023mmt}
T.D.~Brennan and S.~Hong, \emph{{Introduction to Generalized Global Symmetries in QFT and Particle Physics}},  \href{https://arxiv.org/abs/2306.00912}{{\ttfamily 2306.00912}}.

\bibitem{Costa:2024wks}
D.~Costa et~al., \emph{{Simons Lectures on Categorical Symmetries}},  11, 2024 [\href{https://arxiv.org/abs/2411.09082}{{\ttfamily 2411.09082}}].

\bibitem{Bhardwaj:2023kri}
L.~Bhardwaj, L.E.~Bottini, L.~Fraser-Taliente, L.~Gladden, D.S.W.~Gould, A.~Platschorre et~al., \emph{{Lectures on generalized symmetries}}, \href{https://doi.org/10.1016/j.physrep.2023.11.002}{\emph{Phys. Rept.} {\bfseries 1051} (2024) 1} [\href{https://arxiv.org/abs/2307.07547}{{\ttfamily 2307.07547}}].

\bibitem{Sheckler:2025rlk}
A.~Sheckler, \emph{{Mixed Anomalies of Magnetic Symmetries}},  \href{https://arxiv.org/abs/2503.08789}{{\ttfamily 2503.08789}}.

\bibitem{Bhardwaj:2022dyt}
L.~Bhardwaj, M.~Bullimore, A.E.V.~Ferrari and S.~Schafer-Nameki, \emph{{Anomalies of generalized symmetries from solitonic defects}}, \href{https://doi.org/10.21468/SciPostPhys.16.3.087}{\emph{SciPost Phys.} {\bfseries 16} (2024) 087} [\href{https://arxiv.org/abs/2205.15330}{{\ttfamily 2205.15330}}].

\bibitem{Pace:2023kyi}
S.D.~Pace, \emph{{Emergent generalized symmetries in ordered phases and applications to quantum disordering}}, \href{https://doi.org/10.21468/SciPostPhys.17.3.080}{\emph{SciPost Phys.} {\bfseries 17} (2024) 080} [\href{https://arxiv.org/abs/2308.05730}{{\ttfamily 2308.05730}}].

\bibitem{Chen:2022cyw}
S.~Chen and Y.~Tanizaki, \emph{{Solitonic Symmetry beyond Homotopy: Invertibility from Bordism and Noninvertibility from Topological Quantum Field Theory}}, \href{https://doi.org/10.1103/PhysRevLett.131.011602}{\emph{Phys. Rev. Lett.} {\bfseries 131} (2023) 011602} [\href{https://arxiv.org/abs/2210.13780}{{\ttfamily 2210.13780}}].

\bibitem{Chen:2023czk}
S.~Chen and Y.~Tanizaki, \emph{{Solitonic symmetry as non-invertible symmetry: cohomology theories with TQFT coefficients}},  \href{https://arxiv.org/abs/2307.00939}{{\ttfamily 2307.00939}}.

\bibitem{Pace:2023mdo}
S.D.~Pace, C.~Zhu, A.~Beaudry and X.-G.~Wen, \emph{{Generalized symmetries in singularity-free nonlinear {\ensuremath{\sigma}} models and their disordered phases}}, \href{https://doi.org/10.1103/PhysRevB.110.195149}{\emph{Phys. Rev. B} {\bfseries 110} (2024) 195149} [\href{https://arxiv.org/abs/2310.08554}{{\ttfamily 2310.08554}}].

\bibitem{Seiberg:1996bd}
N.~Seiberg, \emph{{Five-dimensional SUSY field theories, nontrivial fixed points and string dynamics}}, \href{https://doi.org/10.1016/S0370-2693(96)01215-4}{\emph{Phys. Lett. B} {\bfseries 388} (1996) 753} [\href{https://arxiv.org/abs/hep-th/9608111}{{\ttfamily hep-th/9608111}}].

\bibitem{Lee:1997vp}
K.-M.~Lee and P.~Yi, \emph{{Monopoles and instantons on partially compactified D-branes}}, \href{https://doi.org/10.1103/PhysRevD.56.3711}{\emph{Phys. Rev. D} {\bfseries 56} (1997) 3711} [\href{https://arxiv.org/abs/hep-th/9702107}{{\ttfamily hep-th/9702107}}].

\bibitem{Lee:1998vu}
K.-M.~Lee, \emph{{Instantons and magnetic monopoles on R**3 x S**1 with arbitrary simple gauge groups}}, \href{https://doi.org/10.1016/S0370-2693(98)00283-4}{\emph{Phys. Lett. B} {\bfseries 426} (1998) 323} [\href{https://arxiv.org/abs/hep-th/9802012}{{\ttfamily hep-th/9802012}}].

\bibitem{Lambert:2010iw}
N.~Lambert, C.~Papageorgakis and M.~Schmidt-Sommerfeld, \emph{{M5-Branes, D4-Branes and Quantum 5D super-Yang-Mills}}, \href{https://doi.org/10.1007/JHEP01(2011)083}{\emph{JHEP} {\bfseries 01} (2011) 083} [\href{https://arxiv.org/abs/1012.2882}{{\ttfamily 1012.2882}}].

\bibitem{Tachikawa:2011ch}
Y.~Tachikawa, \emph{{On S-duality of 5d super Yang-Mills on $S^1$}}, \href{https://doi.org/10.1007/JHEP11(2011)123}{\emph{JHEP} {\bfseries 11} (2011) 123} [\href{https://arxiv.org/abs/1110.0531}{{\ttfamily 1110.0531}}].

\bibitem{Kraan:1998kp}
T.C.~Kraan and P.~van Baal, \emph{{Exact T duality between calorons and Taub - NUT spaces}}, \href{https://doi.org/10.1016/S0370-2693(98)00411-0}{\emph{Phys. Lett. B} {\bfseries 428} (1998) 268} [\href{https://arxiv.org/abs/hep-th/9802049}{{\ttfamily hep-th/9802049}}].

\bibitem{Davies:2000nw}
N.M.~Davies, T.J.~Hollowood and V.V.~Khoze, \emph{{Monopoles, affine algebras and the gluino condensate}}, \href{https://doi.org/10.1063/1.1586477}{\emph{J. Math. Phys.} {\bfseries 44} (2003) 3640} [\href{https://arxiv.org/abs/hep-th/0006011}{{\ttfamily hep-th/0006011}}].

\bibitem{Hanany:2001iy}
A.~Hanany and J.~Troost, \emph{{Orientifold planes, affine algebras and magnetic monopoles}}, \href{https://doi.org/10.1088/1126-6708/2001/08/021}{\emph{JHEP} {\bfseries 08} (2001) 021} [\href{https://arxiv.org/abs/hep-th/0107153}{{\ttfamily hep-th/0107153}}].

\bibitem{Freed:2022qnc}
D.S.~Freed, G.W.~Moore and C.~Teleman, \emph{{Topological symmetry in quantum field theory}},  \href{https://arxiv.org/abs/2209.07471}{{\ttfamily 2209.07471}}.

\bibitem{Bhardwaj:2023ayw}
L.~Bhardwaj and S.~Schafer-Nameki, \emph{{Generalized Charges, Part II: Non-Invertible Symmetries and the Symmetry TFT}},  \href{https://arxiv.org/abs/2305.17159}{{\ttfamily 2305.17159}}.

\bibitem{Kapustin:2014gua}
A.~Kapustin and N.~Seiberg, \emph{{Coupling a QFT to a TQFT and Duality}}, \href{https://doi.org/10.1007/JHEP04(2014)001}{\emph{JHEP} {\bfseries 04} (2014) 001} [\href{https://arxiv.org/abs/1401.0740}{{\ttfamily 1401.0740}}].

\bibitem{Ji:2019jhk}
W.~Ji and X.-G.~Wen, \emph{{Categorical symmetry and noninvertible anomaly in symmetry-breaking and topological phase transitions}}, \href{https://doi.org/10.1103/PhysRevResearch.2.033417}{\emph{Phys. Rev. Res.} {\bfseries 2} (2020) 033417} [\href{https://arxiv.org/abs/1912.13492}{{\ttfamily 1912.13492}}].

\bibitem{Apruzzi:2021nmk}
F.~Apruzzi, F.~Bonetti, I.~Garc{\'\i}a~Etxebarria, S.S.~Hosseini and S.~Schafer-Nameki, \emph{{Symmetry TFTs from String Theory}}, \href{https://doi.org/10.1007/s00220-023-04737-2}{\emph{Commun. Math. Phys.} {\bfseries 402} (2023) 895} [\href{https://arxiv.org/abs/2112.02092}{{\ttfamily 2112.02092}}].

\bibitem{Gaiotto:2020iye}
D.~Gaiotto and J.~Kulp, \emph{{Orbifold groupoids}}, \href{https://doi.org/10.1007/JHEP02(2021)132}{\emph{JHEP} {\bfseries 02} (2021) 132} [\href{https://arxiv.org/abs/2008.05960}{{\ttfamily 2008.05960}}].

\bibitem{Witten:1996hc}
E.~Witten, \emph{{Five-brane effective action in $M$-theory.}}, \href{https://doi.org/10.1016/S0393-0440(97)80160-X}{\emph{J. Geom. Phys.} {\bfseries 22} (1997) 103} [\href{https://arxiv.org/abs/hep-th/9610234}{{\ttfamily hep-th/9610234}}].

\bibitem{Witten:1998wy}
E.~Witten, \emph{{AdS/CFT correspondence and topological field theory.}}, \href{https://doi.org/10.1088/1126-6708/1998/12/012}{\emph{JHEP} {\bfseries 12} (1998) 012} [\href{https://arxiv.org/abs/hep-th/9812012}{{\ttfamily hep-th/9812012}}].

\bibitem{Gukov:2020btk}
S.~Gukov, P.-S.~Hsin and D.~Pei, \emph{{Generalized global symmetries of $T[M]$ theories. Part I}}, \href{https://doi.org/10.1007/JHEP04(2021)232}{\emph{JHEP} {\bfseries 04} (2021) 232} [\href{https://arxiv.org/abs/2010.15890}{{\ttfamily 2010.15890}}].

\bibitem{Bertolini:2025wyj}
M.~Bertolini, L.~Di~Pietro, S.C.~Lanza, P.~Niro and A.~Santaniello, \emph{{Symmetry extension by condensation defects in five-dimensional gauge theories}},  \href{https://arxiv.org/abs/2509.16165}{{\ttfamily 2509.16165}}.

\bibitem{BenettiGenolini:2020doj}
P.~Benetti~Genolini and L.~Tizzano, \emph{{Instantons, symmetries and anomalies in five dimensions}}, \href{https://doi.org/10.1007/JHEP04(2021)188}{\emph{JHEP} {\bfseries 04} (2021) 188} [\href{https://arxiv.org/abs/2009.07873}{{\ttfamily 2009.07873}}].

\bibitem{Garcia-Valdecasas:2024cqn}
E.~Garc{\'\i}a-Valdecasas, M.~Reece and M.~Suzuki, \emph{{Monopole breaking of Chern-Weil symmetries}}, \href{https://doi.org/10.21468/SciPostPhys.18.5.162}{\emph{SciPost Phys.} {\bfseries 18} (2025) 162} [\href{https://arxiv.org/abs/2408.00067}{{\ttfamily 2408.00067}}].

\bibitem{Heidenreich:2020pkc}
B.~Heidenreich, J.~McNamara, M.~Montero, M.~Reece, T.~Rudelius and I.~Valenzuela, \emph{{Chern-Weil global symmetries and how quantum gravity avoids them}}, \href{https://doi.org/10.1007/JHEP11(2021)053}{\emph{JHEP} {\bfseries 11} (2021) 053} [\href{https://arxiv.org/abs/2012.00009}{{\ttfamily 2012.00009}}].

\bibitem{Lambert:2014jna}
N.~Lambert, C.~Papageorgakis and M.~Schmidt-Sommerfeld, \emph{{Instanton Operators in Five-Dimensional Gauge Theories}}, \href{https://doi.org/10.1007/JHEP03(2015)019}{\emph{JHEP} {\bfseries 03} (2015) 019} [\href{https://arxiv.org/abs/1412.2789}{{\ttfamily 1412.2789}}].

\bibitem{HOOFT1974276}
G.~Hooft, \emph{Magnetic monopoles in unified gauge theories}, \href{https://doi.org/https://doi.org/10.1016/0550-3213(74)90486-6}{\emph{Nuclear Physics B} {\bfseries 79} (1974) 276}.

\bibitem{Lambert:1999ua}
N.D.~Lambert and D.~Tong, \emph{{Dyonic instantons in five-dimensional gauge theories}}, \href{https://doi.org/10.1016/S0370-2693(99)00894-1}{\emph{Phys. Lett. B} {\bfseries 462} (1999) 89} [\href{https://arxiv.org/abs/hep-th/9907014}{{\ttfamily hep-th/9907014}}].

\bibitem{GarciaGarcia:2025uub}
I.~Garcia~Garcia, M.~Kongsore and K.~Van~Tilburg, \emph{{Dyon Loops and Abelian Instantons}},  \href{https://arxiv.org/abs/2506.14867}{{\ttfamily 2506.14867}}.

\bibitem{Freed:2017rlk}
D.S.~Freed, Z.~Komargodski and N.~Seiberg, \emph{{The Sum Over Topological Sectors and $\theta$ in the 2+1-Dimensional $\mathbb{C}\mathbb{P}^1$ $\sigma$-Model}}, \href{https://doi.org/10.1007/s00220-018-3093-0}{\emph{Commun. Math. Phys.} {\bfseries 362} (2018) 167} [\href{https://arxiv.org/abs/1707.05448}{{\ttfamily 1707.05448}}].

\bibitem{Hsin:2018vcg}
P.-S.~Hsin, H.T.~Lam and N.~Seiberg, \emph{{Comments on One-Form Global Symmetries and Their Gauging in 3d and 4d}}, \href{https://doi.org/10.21468/SciPostPhys.6.3.039}{\emph{SciPost Phys.} {\bfseries 6} (2019) 039} [\href{https://arxiv.org/abs/1812.04716}{{\ttfamily 1812.04716}}].

\bibitem{Choi:2022jqy}
Y.~Choi, H.T.~Lam and S.-H.~Shao, \emph{{Noninvertible Global Symmetries in the Standard Model}}, \href{https://doi.org/10.1103/PhysRevLett.129.161601}{\emph{Phys. Rev. Lett.} {\bfseries 129} (2022) 161601} [\href{https://arxiv.org/abs/2205.05086}{{\ttfamily 2205.05086}}].

\bibitem{Cordova:2022ieu}
C.~Cordova and K.~Ohmori, \emph{{Noninvertible Chiral Symmetry and Exponential Hierarchies}}, \href{https://doi.org/10.1103/PhysRevX.13.011034}{\emph{Phys. Rev. X} {\bfseries 13} (2023) 011034} [\href{https://arxiv.org/abs/2205.06243}{{\ttfamily 2205.06243}}].

\bibitem{Copetti:2023mcq}
C.~Copetti, M.~Del~Zotto, K.~Ohmori and Y.~Wang, \emph{{Higher Structure of Chiral Symmetry}}, \href{https://doi.org/10.1007/s00220-024-05227-9}{\emph{Commun. Math. Phys.} {\bfseries 406} (2025) 73} [\href{https://arxiv.org/abs/2305.18282}{{\ttfamily 2305.18282}}].

\bibitem{Choi:2021kmx}
Y.~Choi, C.~Cordova, P.-S.~Hsin, H.T.~Lam and S.-H.~Shao, \emph{{Noninvertible duality defects in 3+1 dimensions}}, \href{https://doi.org/10.1103/PhysRevD.105.125016}{\emph{Phys. Rev. D} {\bfseries 105} (2022) 125016} [\href{https://arxiv.org/abs/2111.01139}{{\ttfamily 2111.01139}}].

\bibitem{Kaidi:2021xfk}
J.~Kaidi, K.~Ohmori and Y.~Zheng, \emph{{Kramers-Wannier-like Duality Defects in (3+1)D Gauge Theories}}, \href{https://doi.org/10.1103/PhysRevLett.128.111601}{\emph{Phys. Rev. Lett.} {\bfseries 128} (2022) 111601} [\href{https://arxiv.org/abs/2111.01141}{{\ttfamily 2111.01141}}].

\bibitem{Moore_FK}
G.~Moore, ``Lecture notes for felix klein lectures on n=2 gauge theories.'' \url{https://www.physics.rutgers.edu/~gmoore/FelixKleinLectureNotes.pdf}.

\bibitem{Heckman:2018jxk}
J.J.~Heckman and T.~Rudelius, \emph{{Top Down Approach to 6D SCFTs}}, \href{https://doi.org/10.1088/1751-8121/aafc81}{\emph{J. Phys. A} {\bfseries 52} (2019) 093001} [\href{https://arxiv.org/abs/1805.06467}{{\ttfamily 1805.06467}}].

\bibitem{Witten:1995zh}
E.~Witten, \emph{{Some comments on string dynamics}},  in \emph{{STRINGS 95: Future Perspectives in String Theory}}, pp.~501--523, 7, 1995 [\href{https://arxiv.org/abs/hep-th/9507121}{{\ttfamily hep-th/9507121}}].

\bibitem{Seiberg:1996qx}
N.~Seiberg, \emph{{Nontrivial fixed points of the renormalization group in six-dimensions}}, \href{https://doi.org/10.1016/S0370-2693(96)01424-4}{\emph{Phys. Lett. B} {\bfseries 390} (1997) 169} [\href{https://arxiv.org/abs/hep-th/9609161}{{\ttfamily hep-th/9609161}}].

\bibitem{Strominger:1995ac}
A.~Strominger and M.~Dine, \emph{{Open p-branes}}, \href{https://doi.org/10.1201/9781482268737-13}{\emph{Phys. Lett. B} {\bfseries 383} (1996) 44} [\href{https://arxiv.org/abs/hep-th/9512059}{{\ttfamily hep-th/9512059}}].

\bibitem{Freed:2012bs}
D.S.~Freed and C.~Teleman, \emph{{Relative quantum field theory}}, \href{https://doi.org/10.1007/s00220-013-1880-1}{\emph{Commun. Math. Phys.} {\bfseries 326} (2014) 459} [\href{https://arxiv.org/abs/1212.1692}{{\ttfamily 1212.1692}}].

\bibitem{Tachikawa:2013hya}
Y.~Tachikawa, \emph{{On the 6d origin of discrete additional data of 4d gauge theories}}, \href{https://doi.org/10.1007/JHEP05(2014)020}{\emph{JHEP} {\bfseries 05} (2014) 020} [\href{https://arxiv.org/abs/1309.0697}{{\ttfamily 1309.0697}}].

\bibitem{Harvey:1998bx}
J.A.~Harvey, R.~Minasian and G.W.~Moore, \emph{{NonAbelian tensor multiplet anomalies}}, \href{https://doi.org/10.1088/1126-6708/1998/09/004}{\emph{JHEP} {\bfseries 09} (1998) 004} [\href{https://arxiv.org/abs/hep-th/9808060}{{\ttfamily hep-th/9808060}}].

\bibitem{Intriligator:2000eq}
K.A.~Intriligator, \emph{{Anomaly matching and a Hopf-Wess-Zumino term in 6d, N=(2,0) field theories}}, \href{https://doi.org/10.1016/S0550-3213(00)00148-6}{\emph{Nucl. Phys. B} {\bfseries 581} (2000) 257} [\href{https://arxiv.org/abs/hep-th/0001205}{{\ttfamily hep-th/0001205}}].

\bibitem{Maxfield:2012aw}
T.~Maxfield and S.~Sethi, \emph{{The Conformal Anomaly of M5-Branes}}, \href{https://doi.org/10.1007/JHEP06(2012)075}{\emph{JHEP} {\bfseries 06} (2012) 075} [\href{https://arxiv.org/abs/1204.2002}{{\ttfamily 1204.2002}}].

\bibitem{Ohmori:2014kda}
K.~Ohmori, H.~Shimizu, Y.~Tachikawa and K.~Yonekura, \emph{{Anomaly polynomial of general 6d SCFTs}}, \href{https://doi.org/10.1093/ptep/ptu140}{\emph{PTEP} {\bfseries 2014} (2014) 103B07} [\href{https://arxiv.org/abs/1408.5572}{{\ttfamily 1408.5572}}].

\bibitem{Monnier:2017klz}
S.~Monnier, \emph{{The anomaly field theories of six-dimensional (2,0) superconformal theories}}, \href{https://doi.org/10.4310/ATMP.2018.v22.n8.a6}{\emph{Adv. Theor. Math. Phys.} {\bfseries 22} (2018) 2035} [\href{https://arxiv.org/abs/1706.01903}{{\ttfamily 1706.01903}}].

\bibitem{Monnier:2014txa}
S.~Monnier, \emph{{The global anomalies of (2,0) superconformal field theories in six dimensions}}, \href{https://doi.org/10.1007/JHEP09(2014)088}{\emph{JHEP} {\bfseries 09} (2014) 088} [\href{https://arxiv.org/abs/1406.4540}{{\ttfamily 1406.4540}}].

\bibitem{Sheckler:2025lfv}
A.~Sheckler, \emph{{Anomaly Matching in 6d $\mathcal{N}=(2,0)$ SCFTs from M5 Cobordism}},  \href{https://arxiv.org/abs/2501.04785}{{\ttfamily 2501.04785}}.

\bibitem{Monnier:2016jlo}
S.~Monnier, \emph{{Topological field theories on manifolds with Wu structures}}, \href{https://doi.org/10.1142/S0129055X17500155}{\emph{Rev. Math. Phys.} {\bfseries 29} (2017) 1750015} [\href{https://arxiv.org/abs/1607.01396}{{\ttfamily 1607.01396}}].

\bibitem{DelZotto:2015isa}
M.~Del~Zotto, J.J.~Heckman, D.S.~Park and T.~Rudelius, \emph{{On the Defect Group of a 6D SCFT}}, \href{https://doi.org/10.1007/s11005-016-0839-5}{\emph{Lett. Math. Phys.} {\bfseries 106} (2016) 765} [\href{https://arxiv.org/abs/1503.04806}{{\ttfamily 1503.04806}}].

\bibitem{Bhardwaj:2020phs}
L.~Bhardwaj and S.~Sch{\"a}fer-Nameki, \emph{{Higher-form symmetries of 6d and 5d theories}}, \href{https://doi.org/10.1007/JHEP02(2021)159}{\emph{JHEP} {\bfseries 02} (2021) 159} [\href{https://arxiv.org/abs/2008.09600}{{\ttfamily 2008.09600}}].

\bibitem{Apruzzi:2022dlm}
F.~Apruzzi, \emph{{Higher form symmetries TFT in 6d}}, \href{https://doi.org/10.1007/JHEP11(2022)050}{\emph{JHEP} {\bfseries 11} (2022) 050} [\href{https://arxiv.org/abs/2203.10063}{{\ttfamily 2203.10063}}].

\bibitem{Cordova:2020tij}
C.~Cordova, T.T.~Dumitrescu and K.~Intriligator, \emph{{2-Group Global Symmetries and Anomalies in Six-Dimensional Quantum Field Theories}}, \href{https://doi.org/10.1007/JHEP04(2021)252}{\emph{JHEP} {\bfseries 04} (2021) 252} [\href{https://arxiv.org/abs/2009.00138}{{\ttfamily 2009.00138}}].

\bibitem{Apruzzi:2021vcu}
F.~Apruzzi, S.~Schafer-Nameki, L.~Bhardwaj and J.~Oh, \emph{{The Global Form of Flavor Symmetries and 2-Group Symmetries in 5d SCFTs}}, \href{https://doi.org/10.21468/SciPostPhys.13.2.024}{\emph{SciPost Phys.} {\bfseries 13} (2022) 024} [\href{https://arxiv.org/abs/2105.08724}{{\ttfamily 2105.08724}}].

\bibitem{Apruzzi:2020zot}
F.~Apruzzi, M.~Dierigl and L.~Lin, \emph{{The fate of discrete 1-form symmetries in 6d}}, \href{https://doi.org/10.21468/SciPostPhys.12.2.047}{\emph{SciPost Phys.} {\bfseries 12} (2022) 047} [\href{https://arxiv.org/abs/2008.09117}{{\ttfamily 2008.09117}}].

\bibitem{Apruzzi:2024cty}
F.~Apruzzi, S.~Schafer-Nameki and A.~Warman, \emph{{Non-Invertible Symmetries in 6d from Green-Schwarz Automorphisms}},  \href{https://arxiv.org/abs/2411.09674}{{\ttfamily 2411.09674}}.

\bibitem{Bonetti:2024etn}
F.~Bonetti, M.~Del~Zotto and R.~Minasian, \emph{{SymTFTs and non-invertible symmetries of 6d (2,0) SCFTs of type D from M-theory}}, \href{https://doi.org/10.1007/JHEP02(2025)156}{\emph{JHEP} {\bfseries 02} (2025) 156} [\href{https://arxiv.org/abs/2412.07842}{{\ttfamily 2412.07842}}].

\bibitem{Douglas:2010iu}
M.R.~Douglas, \emph{{On D=5 super Yang-Mills theory and (2,0) theory}}, \href{https://doi.org/10.1007/JHEP02(2011)011}{\emph{JHEP} {\bfseries 02} (2011) 011} [\href{https://arxiv.org/abs/1012.2880}{{\ttfamily 1012.2880}}].

\bibitem{Kim:2011mv}
H.-C.~Kim, S.~Kim, E.~Koh, K.~Lee and S.~Lee, \emph{{On instantons as Kaluza-Klein modes of M5-branes}}, \href{https://doi.org/10.1007/JHEP12(2011)031}{\emph{JHEP} {\bfseries 12} (2011) 031} [\href{https://arxiv.org/abs/1110.2175}{{\ttfamily 1110.2175}}].

\bibitem{Cordova:2015vwa}
C.~Cordova, T.T.~Dumitrescu and X.~Yin, \emph{{Higher derivative terms, toroidal compactification, and Weyl anomalies in six-dimensional (2, 0) theories}}, \href{https://doi.org/10.1007/JHEP10(2019)128}{\emph{JHEP} {\bfseries 10} (2019) 128} [\href{https://arxiv.org/abs/1505.03850}{{\ttfamily 1505.03850}}].

\bibitem{Duan:2021ges}
Z.~Duan, K.~Lee, J.~Nahmgoong and X.~Wang, \emph{{Twisted 6d (2, 0) SCFTs on a circle}}, \href{https://doi.org/10.1007/JHEP07(2021)179}{\emph{JHEP} {\bfseries 07} (2021) 179} [\href{https://arxiv.org/abs/2103.06044}{{\ttfamily 2103.06044}}].

\bibitem{Lambert:2010wm}
N.~Lambert and C.~Papageorgakis, \emph{{Nonabelian (2,0) Tensor Multiplets and 3-algebras}}, \href{https://doi.org/10.1007/JHEP08(2010)083}{\emph{JHEP} {\bfseries 08} (2010) 083} [\href{https://arxiv.org/abs/1007.2982}{{\ttfamily 1007.2982}}].

\bibitem{Paban:1998ea}
S.~Paban, S.~Sethi and M.~Stern, \emph{{Constraints from extended supersymmetry in quantum mechanics}}, \href{https://doi.org/10.1016/S0550-3213(98)00518-5}{\emph{Nucl. Phys. B} {\bfseries 534} (1998) 137} [\href{https://arxiv.org/abs/hep-th/9805018}{{\ttfamily hep-th/9805018}}].

\bibitem{Paban:1998qy}
S.~Paban, S.~Sethi and M.~Stern, \emph{{Supersymmetry and higher derivative terms in the effective action of Yang-Mills theories}}, \href{https://doi.org/10.1088/1126-6708/1998/06/012}{\emph{JHEP} {\bfseries 06} (1998) 012} [\href{https://arxiv.org/abs/hep-th/9806028}{{\ttfamily hep-th/9806028}}].

\bibitem{Sethi:1999qv}
S.~Sethi and M.~Stern, \emph{{Supersymmetry and the Yang-Mills effective action at finite N}}, \href{https://doi.org/10.1088/1126-6708/1999/06/004}{\emph{JHEP} {\bfseries 06} (1999) 004} [\href{https://arxiv.org/abs/hep-th/9903049}{{\ttfamily hep-th/9903049}}].

\bibitem{Movshev:2009ba}
M.~Movshev and A.~Schwarz, \emph{{Supersymmetric Deformations of Maximally Supersymmetric Gauge Theories}}, \href{https://doi.org/10.1007/JHEP09(2012)136}{\emph{JHEP} {\bfseries 09} (2012) 136} [\href{https://arxiv.org/abs/0910.0620}{{\ttfamily 0910.0620}}].

\bibitem{Lawrie:2023tdz}
C.~Lawrie, X.~Yu and H.Y.~Zhang, \emph{{Intermediate defect groups, polarization pairs, and noninvertible duality defects}}, \href{https://doi.org/10.1103/PhysRevD.109.026005}{\emph{Phys. Rev. D} {\bfseries 109} (2024) 026005} [\href{https://arxiv.org/abs/2306.11783}{{\ttfamily 2306.11783}}].

\bibitem{Bashmakov:2022uek}
V.~Bashmakov, M.~Del~Zotto, A.~Hasan and J.~Kaidi, \emph{{Non-invertible symmetries of class S theories}}, \href{https://doi.org/10.1007/JHEP05(2023)225}{\emph{JHEP} {\bfseries 05} (2023) 225} [\href{https://arxiv.org/abs/2211.05138}{{\ttfamily 2211.05138}}].

\bibitem{Cordova:2018cvg}
C.~C{\'o}rdova, T.T.~Dumitrescu and K.~Intriligator, \emph{{Exploring 2-Group Global Symmetries}}, \href{https://doi.org/10.1007/JHEP02(2019)184}{\emph{JHEP} {\bfseries 02} (2019) 184} [\href{https://arxiv.org/abs/1802.04790}{{\ttfamily 1802.04790}}].

\bibitem{Benini:2018reh}
F.~Benini, C.~C{\'o}rdova and P.-S.~Hsin, \emph{{On 2-Group Global Symmetries and their Anomalies}}, \href{https://doi.org/10.1007/JHEP03(2019)118}{\emph{JHEP} {\bfseries 03} (2019) 118} [\href{https://arxiv.org/abs/1803.09336}{{\ttfamily 1803.09336}}].

\bibitem{Brennan:2024fgj}
T.D.~Brennan and Z.~Sun, \emph{{A SymTFT for continuous symmetries}}, \href{https://doi.org/10.1007/JHEP12(2024)100}{\emph{JHEP} {\bfseries 12} (2024) 100} [\href{https://arxiv.org/abs/2401.06128}{{\ttfamily 2401.06128}}].

\bibitem{Antinucci:2024zjp}
A.~Antinucci and F.~Benini, \emph{{Anomalies and gauging of U(1) symmetries}}, \href{https://doi.org/10.1103/PhysRevB.111.024110}{\emph{Phys. Rev. B} {\bfseries 111} (2025) 024110} [\href{https://arxiv.org/abs/2401.10165}{{\ttfamily 2401.10165}}].

\bibitem{Aharony:2013hda}
O.~Aharony, N.~Seiberg and Y.~Tachikawa, \emph{{Reading between the lines of four-dimensional gauge theories}}, \href{https://doi.org/10.1007/JHEP08(2013)115}{\emph{JHEP} {\bfseries 08} (2013) 115} [\href{https://arxiv.org/abs/1305.0318}{{\ttfamily 1305.0318}}].

\bibitem{Kapustin:2013qsa}
A.~Kapustin and R.~Thorngren, \emph{{Topological Field Theory on a Lattice, Discrete Theta-Angles and Confinement}}, \href{https://doi.org/10.4310/ATMP.2014.v18.n5.a4}{\emph{Adv. Theor. Math. Phys.} {\bfseries 18} (2014) 1233} [\href{https://arxiv.org/abs/1308.2926}{{\ttfamily 1308.2926}}].

\bibitem{Bolognesi:2011nh}
S.~Bolognesi and K.~Lee, \emph{{Instanton Partons in 5-dim SU(N) Gauge Theory}}, \href{https://doi.org/10.1103/PhysRevD.84.106001}{\emph{Phys. Rev. D} {\bfseries 84} (2011) 106001} [\href{https://arxiv.org/abs/1106.3664}{{\ttfamily 1106.3664}}].

\bibitem{Collie:2009iz}
B.~Collie and D.~Tong, \emph{{The Partonic Nature of Instantons}}, \href{https://doi.org/10.1088/1126-6708/2009/08/006}{\emph{JHEP} {\bfseries 08} (2009) 006} [\href{https://arxiv.org/abs/0905.2267}{{\ttfamily 0905.2267}}].

\bibitem{Boyarsky:2002ck}
A.~Boyarsky, J.A.~Harvey and O.~Ruchayskiy, \emph{{A Toy model of the M5-brane: Anomalies of monopole strings in five dimensions}}, \href{https://doi.org/10.1006/aphy.2002.6294}{\emph{Annals Phys.} {\bfseries 301} (2002) 1} [\href{https://arxiv.org/abs/hep-th/0203154}{{\ttfamily hep-th/0203154}}].

\bibitem{Apruzzi:2025hvs}
F.~Apruzzi, N.~Dondi, I.~Garc{\'\i}a~Etxebarria, H.T.~Lam and S.~Schafer-Nameki, \emph{{Symmetry TFTs for Continuous Spacetime Symmetries}},  \href{https://arxiv.org/abs/2509.07965}{{\ttfamily 2509.07965}}.

\end{thebibliography}\endgroup

\end{document}